\documentclass{jfm}
\usepackage{graphicx}
\usepackage{newtxtext}
\usepackage{newtxmath}
\usepackage{natbib}
\usepackage{caption}
\usepackage{hyperref}
\usepackage{bm}
\usepackage[dvipsnames]{xcolor}
\hypersetup{
    colorlinks = true,
    urlcolor   = blue,
    citecolor  = CadetBlue,
}

\title{Convective mesoscale turbulence at very low Prandtl numbers}

\author{Ambrish Pandey\aff{1}
        \corresp{\email{ambrish.pandey@nyu.edu}},
        Dmitry Krasnov\aff{2},
        Katepalli R. Sreenivasan\aff{1,3,4},
        \and J\"org Schumacher\aff{2,3}
        \corresp{\email{joerg.schumacher@tu-ilmenau.de}}}

\affiliation{\aff{1}Center for Space Science, New York University Abu Dhabi, PO Box 129188 Abu Dhabi, UAE
\aff{2}Institut f\"ur Thermo- und Fluiddynamik, Technische Universit\"at Ilmenau, PO Box 100565, D-98684 Ilmenau, Germany
\aff{3} Tandon School of Engineering, New York University, New York NY 11201, USA
\aff{4} Department of Physics and Courant Institute of Mathematical Sciences, New York University, New York, NY 11201, USA
}

\begin{document}
\maketitle

\begin{abstract}
Horizontally extended turbulent convection, termed mesoscale convection in natural systems, remains a challenge to investigate in both experiments and simulations. This is particularly so for very low molecular Prandtl numbers as in stellar convection and in Earth's outer core. The present study reports three-dimensional direct numerical simulations of turbulent Rayleigh-B\'{e}nard convection in square boxes of side length $L$ and height $H$ with the aspect ratio $\Gamma =L/H$ of 25, for Prandtl numbers that span almost 4 orders of magnitude, $10^{-3}\le Pr \le 7$, and Rayleigh numbers $10^5 \le Ra \le 10^7$, obtained by massively parallel computations on grids of up to $5.36\times 10^{11}$ points. The low end of this $Pr$-range cannot be accessed in controlled laboratory measurements. We report the essential properties of the flow and their trends with Rayleigh and Prandtl numbers, in particular the global transport of momentum and heat---the latter decomposed into convective and diffusive contributions---across the convection layer, mean vertical profiles of the temperature and temperature fluctuations, and the kinetic energy and thermal dissipation rates. We also explore the degree to which the turbulence in the bulk of the convection layer resembles classical homogeneous and isotropic turbulence in terms of spectra, increment moments, and dissipative anomaly, and find close similarities. Finally, we show that a characteristic scale on the order of the mesoscale seems to saturate to a wavelength of $\lambda\gtrsim 3H$ for $Pr\lesssim 0.005$. We briefly discuss possible implications of these results for the development of subgrid scale parameterization of turbulent convection. 
\end{abstract}


\section{Introduction}
\label{sec:intro}

Thermal convection in stellar and planetary interiors and their atmospheres is complex because it is driven by several factors in combination, but researchers often study the idealized model of convection---the Rayleigh-B\'enard convection (RBC)---where a fluid layer bounded between two horizontal plates is heated and cooled uniformly from the bottom and the top, respectively \citep{Tritton:book1977, Siggia:ARFM1994, Kadanoff:PT2001, Ahlers:RMP2009, Chilla:EPJE2012}. An important parameter governing RBC is the Prandtl number $Pr$, which is the ratio of the kinematic viscosity $\nu$ and the thermal diffusivity $\kappa$ of the fluid. Stellar and planetary convection is often characterized by very small Prandtl numbers; for example, $Pr \approx 10^{-6}$ in the Sun \citep{Brandenburg:PR2005,Schumacher:RMP2020,Garaud:PRF2021} and $Pr \approx 10^{-2}-10^{-1}$ in the Earth's outer core~\citep{Calkins:EPSL2012, Aurnou:PEPI2015, Guervilly:Nature2019}. The Rayleigh number $Ra$, which characterises the strength of the driving buoyancy force relative to viscous and thermal dissipative forces, is another control parameter, and is very high in most natural flows. The ratio of the horizontal and vertical extents of a convective flow is the aspect ratio $\Gamma$; a crucial feature of all natural convective flows is that $\Gamma \gg 1$, which allows the formation of turbulent superstructures---coherent flow patterns with characteristic scale larger than the depth of the convection layer~\citep{Cattaneo:APJ2001, Rincon:AA2005}. A possible example is supergranulation on Sun's surface~\citep{Nordlung:LRSP2009,Rincon:LRSP2018}. Even though RBC incorporates a number of simplifications such as the Boussinesq approximation \citep{Tritton:book1977,Schumacher:RMP2020}, it appears that a study of the flow in extended layers at low Prandtl numbers is highly worthwhile. This is the primary objective of the current work. We are also motivated by the relevance of low-$Pr$ convection for industrial applications that use liquid metals.

Despite this relevance, low-$Pr$ turbulent convection has not been explored extensively in experiments mainly because liquid metals such as mercury, gallium, and sodium are difficult to handle and optically opaque~\citep{Cioni:JFM1997, Zuerner:JFM2019}. Even when these difficulties are circumvented, the lowest Prandtl number that can be explored is $Pr \approx 0.006$ for liquid sodium \citep{Horanyi:IJHMT1999}, which is still some three orders of magnitude higher than in the solar convection zone. Direct Numerical Simulations (DNS) offer important tools but they, too, are hindered by demanding resolution requirements due to the highly inertial nature of low-$Pr$ convection~\citep{Breuer:PRE2004,Schumacher:PNAS2015,Scheel:JFM2016,Scheel:PRF2017,Pandey:Nature2018, Zwirner:JFM2020}. The fact that the required computational power increases with increasing aspect ratio as $\Gamma^2$ further limits numerical investigations. Yet, \citet{Pandey:Nature2018} were able to perform DNS of RBC in a $\Gamma = 25$ domain by achieving $Pr$ as low as 0.005 at $Ra = 10^5$ to study turbulent superstructures. In the present work, we significantly extend the parameter range from \citet{Pandey:Nature2018} and \citet{Fonda:PNAS2019} by further decreasing the Prandtl number five-fold, while also increasing the Rayleigh number by two orders of magnitude. The highest Reynolds number achieved in the present work is nearly $5.6 \times 10^4$, requiring massively parallel DNS on computational grids of more than $5\times 10^{11}$ points. 

This work has three main goals: (1) report trends of heat and momentum transfer with respect to Prandtl number over nearly 4 orders of magnitude; (2) assess the closeness of small-scale statistical properties in the bulk of the convection layer to the classical Kolmogorov-type behaviour \citep{Kolmogorov:DANS1941a,Frisch:book}. This assessment comprises energy spectra, a test of the 4/5-th law and an investigation of the dissipative anomaly \citep{Sreenivasan:POF1984, Sreenivasan:POF1998} in turbulent convection flow with boundaries; (3) analyse the large-scale circulation patterns, the turbulent superstructures of convection, particularly their trends with decreasing $Pr$. Note that the convective flows in the Earth's and stellar interiors are also associated with (differential) rotation and magnetic fields that can lead to strong departures from local isotropy at larger and intermediate scales~\citep{Aurnou:PEPI2015}, but we shall here focus on the influence of low Prandtl number. These DNS series will thus provide a unique data base for the parametrization of turbulent transport in mesoscale configurations characterized by a degree of large-scale order, with well-resolved thermal and kinetic energy dissipation rates. 

It is becoming increasingly clear that many properties of low-$Pr$ convective flows differ from those at moderate and high Prandtl numbers. For example, the efficacy of low-$Pr$ flows in transporting heat is lower, and that in transporting momentum higher, than in high-$Pr$ flows; the disparity between the two increases as $Pr$ is lowered~\citep{Scheel:PRF2017}. Due to high (low) thermal (momentum) diffusivity, low-$Pr$ convection exhibits coarser thermal structures but the length scales in the velocity field have a broader distribution. This results in an enhanced separation between the energy injection and energy dissipation scales in low-$Pr$ convective flows~\citep{Schumacher:PNAS2015}. In this regime, the kinetic energy spectrum has been observed to approximate the classical Kolmogorov scaling \citep{Kolmogorov:DANS1941a} with the $-5/3$ power in the inertial range~\citep{Mishra:PRE2010, Lohse:ARFM2010, Bhattacharya:PRF2021}. Here, we analyse the kinetic energy spectra in the bulk region of the flow and show that it indeed exhibits the classical Kolmogorov scaling with an inertial range, showing no tendency towards Bolgiano scaling \citep{Bolgiano:JGR1959}, according to which, the conversion of the kinetic to potential energy leads to the steeper $k^{-11/5}$ scaling in the inertial range~\citep{Lohse:ARFM2010, Verma:NJP2017, Verma:book2018}.

Turbulent superstructures of convection can be characterized by a typical spatial scale $\lambda$ and a temporal scale $\tau$; finer scales evolve much faster than $\tau$. Thus, the scales $\lambda$ and $\tau$ help distinguish the coarse and gradually evolving large-scale patterns from the finer (and faster) turbulent fluctuations~\citep{Pandey:Nature2018, Krug:JFM2020}. The characteristic scales of superstructures have been observed to depend on $Pr$ and $Ra$~\citep{Hartlep:PRL2003, Hartlep:JFM2005, Hardenberg:PLA2008, BailonCuba:JFM2010, Emran:JFM2015, Pandey:Nature2018, Stevens:PRF2018, Schneide:PRF2018, Fonda:PNAS2019, Green:JFM2020, Krug:JFM2020, Pandey:APJ2021, Lenzi:PRE2021}, as well as on the thermal boundary conditions at the horizontal top and bottom plates~\citep{Vieweg:PRR2021}. The characteristic length scale of superstructures, which is nearly twice the depth $H$ of the convection layer at the onset of convection, increases with increasing $Ra$~\citep{Stevens:PRF2018, Pandey:Nature2018}. This dependence on $Pr$ is complex, with $\lambda(Pr)$ showing a peak near $Pr \approx 7$ and decreasing as $Pr$ departs from this value~\citep{Pandey:Nature2018}. This decreasing trend of $\lambda(Pr)$ continues to hold up to a $Pr \approx 0.005$ below which the scales seem to level off at a wavelength of $\lambda\gtrsim 3H$. 

The remainder of this article is organized as follows. In \S~\ref{sec:numerical}, we briefly describe the DNS and note the parameter space explored. In \S~\ref{sec:integral}, we discuss the flow structures and the scaling of the global transport of heat and momentum, and study in \S~\ref{sec:mean} the vertical profiles of temperature, convective and diffusive transports, as well as dissipation rates. Statistical properties of the flow such as kinetic energy spectra and third-order structure function are examined in \S~\ref{sec:bulk} for their compatibility with Kolmogorov forms. The characterisation of turbulent superstructures is presented in \S~\ref{sec:TSS}. We conclude the main findings in \S~\ref{sec:concl}, and present an outlook. Appendices A and B deal with specific tests of sufficient resolution. Appendix C discusses technical detail of the estimates of length and velocity scales of superstructures. 

\section{Details of direct numerical simulations}
\label{sec:numerical}

We perform direct numerical simulations (DNS) of RBC in a closed rectangular domain with square cross-section of length $L = 25H$, where $H$ is the depth of the convection layer. We solve the following non-dimensionalized equations incorporating the Oberbeck-Boussinesq (OB) approximation:
\begin{eqnarray}
\frac{\partial {\bm u}}{\partial t} + ({\bm u} \cdot {\bm \nabla}) {\bm u} & = & -{\bm \nabla}p + T \hat{{\bm z}} + \sqrt{\frac{Pr}{Ra}} \, \nabla^2 {\bm u}, \label{eq:u} \\
\frac{\partial T}{\partial t} + ({\bm u} \cdot {\bm \nabla}) T & = & \frac{1}{\sqrt{PrRa}} \, \nabla^2 T, \label{eq:T} \\
{\bm \nabla} \cdot {\bm u} & = & 0. \label{eq:m}
\end{eqnarray}
Here ${\bm u} \equiv (u_x,u_y,u_z)$, $T$, and $p$ are the velocity, temperature and pressure fields, respectively. The Prandtl number $Pr$ and the Rayleigh number $Ra = \alpha g \Delta T H^3/\nu\kappa$, where $\alpha$ is the coefficient of thermal expansion of the fluid, $g$ is the acceleration due to gravity, and $\Delta T$ is the imposed temperature difference between the horizontal plates. We have used the layer's depth $H$, the free-fall velocity $u_f = \sqrt{\alpha g \Delta T H}$, the free-fall time $t_f = H/u_f$, and $\Delta T$ as the non-dimensionalizing length, velocity, time, and temperature scales, respectively. We use the no-slip condition on all boundaries. We employ the isothermal condition on the horizontal plates and the adiabatic condition on the sidewalls. This allows direct comparison with other simulations at higher $Pr$ \citep{Pandey:Nature2018,Fonda:PNAS2019} and controlled laboratory experiments, such as those of \citet{Moller:2022}, in exactly the same setting to enable a consistent analysis across the $Ra$--$Pr$ parameter plane. 

We use two different solvers for the simulations. For moderate and large $Pr$, we use a spectral element solver {\sc Nek5000}~\citep{Fischer:JCP1997}, where the flow domain is divided into a finite number of elements $N_e$. The Lagrangian interpolation polynomials of order $N$ are further used to expand the turbulence fields within each element~\citep{Scheel:NJP2013}, thus resulting in a total of $N_e N^3$ mesh cells in the entire flow domain. As low-$Pr$ convection is dominated by inertial forces, the resulting flow acquires increased fine structure whose resolution requires more extensive computational resources~\citep{Schumacher:PNAS2015}. Therefore, we performed those simulations using a second-order finite difference solver that requires significantly less working memory at a given grid size. Here, the flow domain is divided into $N_x \times N_y \times N_z$ non-uniform mesh cells~\citep{Krasnov:CF2011, Liu:JFM2018}. We have verified that the results obtained from both the the solvers agree well with each other by performing two simulations for $Pr = 0.005, Ra = 10^5$ and $Pr = 0.7, Ra = 10^7$ using both solvers. We refer to Appendix~\ref{sec:app1} for a direct comparison of globally-averaged and horizontally-averaged convective heat fluxes and dissipation rates from the two solvers. Important parameters of all the simulations are provided in table~\ref{table:details}.

\begin{table}
  \begin{center}
\def~{\hphantom{0}}
  \begin{tabular}{lcccccccc}
$Pr$ & $Ra$ & No. of mesh cells  & $N_{\rm tot} (\rm bn)$ & $Nu$ & $Re$ & $u_\mathrm{rms}$ & $\lambda/H$ & $\tau \, (t_f)$ \\
0.001 & $10^5$ & $9600 \times 9600  \times 640$ & 59 & $1.21 \pm 0.005$ & $4800 \pm 30$ & $0.480 \pm 0.003$ & 3.12 & 25 \\ 
0.005$^\dagger$ & $10^5$ & $2367488 \times 11^3$ & 3.1 & $1.90 \pm 0.01$ & $2491 \pm 20$ & $0.557 \pm 0.001$ & 3.13 & 22 \\
0.021$^\dagger$ & $10^5$ &  $2367488 \times 7^3$ & 0.81 & $2.60 \pm 0.01$ & $1120 \pm 8$ & $0.513 \pm 0.001$ & 3.57 & 26 \\
0.7$^\dagger$ & $10^5$ &  $1352000 \times 5^3$ & 0.17 & $4.26 \pm 0.02$ & $92.0 \pm 0.4$ & $0.243 \pm 0.001$ & 4.14 & 59 \\
7.0$^\dagger$ & $10^5$ &  $1352000 \times 5^3$ & 0.17 & $4.14 \pm 0.01$ & $10.7 \pm 0.03$ & $0.089 \pm 0.001$ & 6.14 & 214 \\
0.001 & $10^6$ & $12800  \times 12800 \times 800$ & 131 & $2.48 \pm 0.005$ & $19876 \pm 1$ & $0.628 \pm 0.0001$ & 3.12 & 19 \\
0.005 & $10^6$ & $8192  \times 8192 \times 512$ & 34 & $3.52 \pm 0.03$ & $7603 \pm 20$ & $0.537 \pm 0.001$ & 3.57 & 24 \\
0.021 & $10^6$ & $8192  \times 8192 \times 512$ & 34 & $4.84 \pm 0.01$ & $3157 \pm 12$ & $0.457 \pm 0.002$ & 4.17 & 32 \\
0.7$^\dagger$ & $10^6$ & $2367488 \times 7^3$ & 0.81 & $8.10 \pm 0.03$ & $290 \pm 1$ & $0.242 \pm 0.001$ & 4.90 & 67 \\
7.0$^{\dagger\dagger}$ & $10^6$ & $2367488 \times 7^3$ & 0.81 & $8.30 \pm 0.02$ & $38.7 \pm 0.2$ & $0.102 \pm 0.001$ & 6.14 & 187 \\
0.001 & $10^7$ & $20480 \times 20480 \times 1280$ & 537 & $4.57 \pm 0.01$ & $56256 \pm 16$ & $0.562 \pm 0.0002$ & 3.57 & 25 \\
0.7$^\dagger$ & $10^7$ & $2367488 \times 11^3$ & 3.1 & $15.48 \pm 0.06$ & $864 \pm 3$ & $0.228 \pm 0.001$ & 6.3 & 72 \\
7.0$^{\dagger\dagger}$ & $10^7$ & $1352000 \times 11^3$ & 1.8 & $16.25 \pm 0.03$ & $130 \pm 0.5$ & $0.109 \pm 0.001$ & 6.2 & 138 
  \end{tabular}
  \caption{Important parameters of the simulations in a rectangular box of $\Gamma = 25$ with square cross-section; the number of mesh cells, $N_x \times N_y \times N_z$ and $N_e \times N^3$, are for the finite difference and spectral element solvers, respectively; $N_{\rm tot}$ represents the total number of mesh cells in units of a billion; $Nu$ and $Re$ are the volume and time averaged Nusselt and Reynolds numbers, respectively; $u_\mathrm{rms}$ is the root-mean-square velocity computed over the entire volume; $\lambda$ and $\tau$ are, respectively, the characteristic length and time scales of turbulent superstructures. Runs with superscript $\dagger$ are taken from \citet{Pandey:Nature2018}, while those with superscript ${\dagger\dagger}$ are taken from \citet{Fonda:PNAS2019}.}
  \label{table:details}
  \end{center}
\end{table}

\section{Flow morphology and global transport}
\label{sec:integral}

\subsection{Structures of velocity and temperature fields}

\begin{figure}
\centerline{\includegraphics[width=1\textwidth]{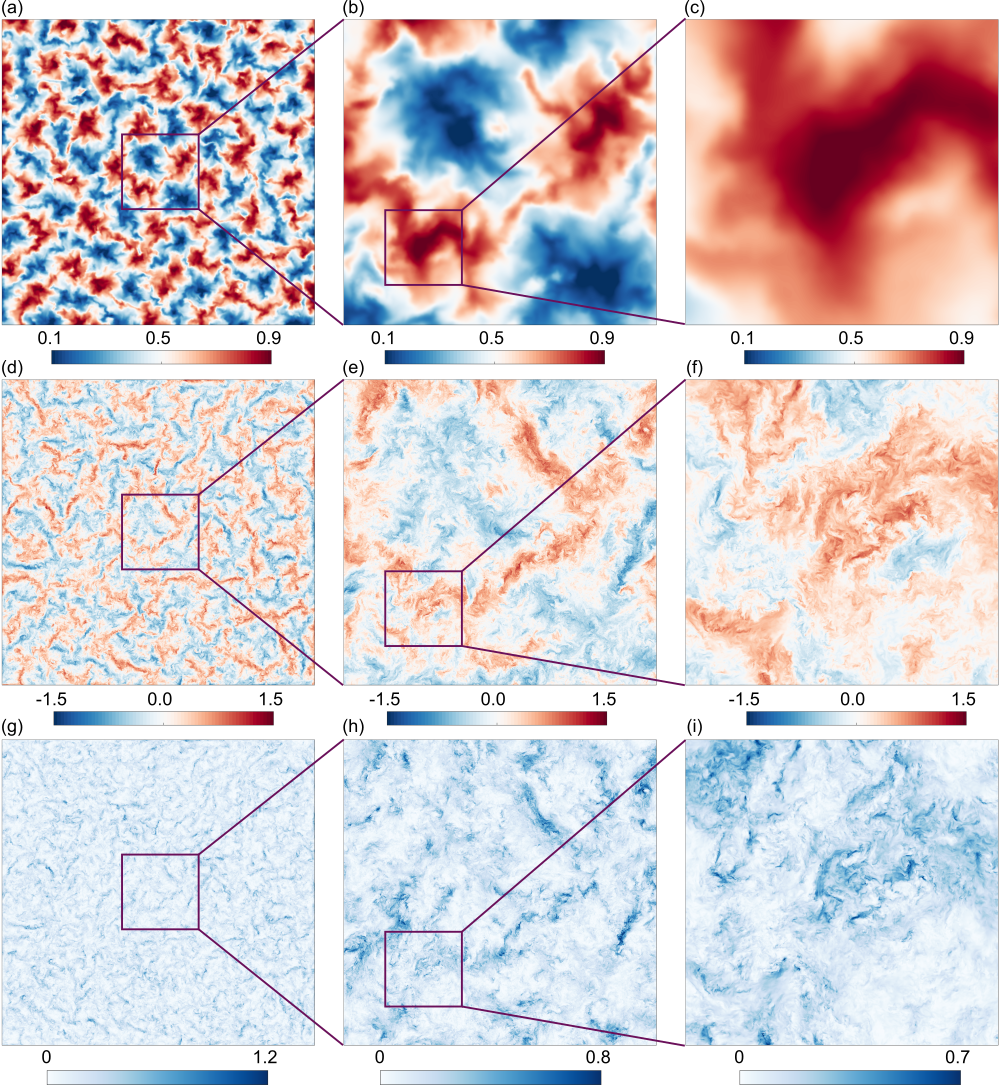}}
\caption{Turbulent superstructures of convection in a low-$Pr$ flow with $Pr = 0.001$ and $Ra = 10^7$. The panels represent instantaneous temperature fields (top row), vertical velocity (middle row), and turbulent kinetic energy in the midplane. In this low-$Pr$ flow, the thermal energy is primarily contained in large-scales, whereas the kinetic energy is distributed over a broad range of scales. Left, middle, and right columns represent fields of view that are $25H \times 25H$, $6.25H \times 6.25H$, and $1.56H \times 1.56H$, respectively.}
\label{fig:T_uz_ke}
\end{figure}

Because the time scales of heat and momentum diffusion processes are very different in low-$Pr$ convection, the temperature field shows coarser structures than the velocity field. This is illustrated in figure~\ref{fig:T_uz_ke}, which displays the instantaneous temperature, vertical velocity, and local turbulent kinetic energy fields in the mid-horizontal plane, $z = H/2$, for the biggest simulations with $Pr = 0.001, Ra = 10^7$. The left panels show the fields in the entire cross-section, and the middle and right panels depict the marked magnifications to highlight small-scale structures. The flow pattern of the temperature and vertical velocity fields show similarities at large scales, but the velocity field also consists of very fine structures compared to the highly diffusive temperature field. The finest scale of the turbulent velocity field, denoted as the Kolmogorov scale $\eta$, is estimated as $\eta = (\nu^3/\langle\varepsilon_u\rangle_{V,t})^{1/4}$. Here, $\langle\varepsilon_u\rangle_{V,t}$ is the combined volume-time average of the kinetic energy dissipation rate field per unit mass, computed at each point by 
\begin{equation}
\varepsilon_u({\bm x},t) = \frac{\nu}{2} \sum_{i,j=1}^3\left( \frac{\partial u_i}{\partial x_j} + \frac{\partial u_j}{\partial x_i} \right)^2, \label{eq:diss_u}
\end{equation}
with $u_i$ representing the velocity component in the direction of coordinate $x_i$. The finest scale of the temperature field is either the Corrsin scale $\eta_C=\eta/Pr^{3/4}$, which marks the end of the inertial-convective range for $Pr<1$, or the Batchelor scale $\eta_B=\eta/Pr^{1/2}$, which marks the end of the viscous-convective range for $Pr>1$; see e.g., \citet{SreenivasanSchumacherPTRSA2010}. It is clear that $\eta < \eta_C$ when $Pr < 1$, and $\eta_B < \eta$ when $Pr>1$. Thus, the finest scales in the flow at hand are either $\eta$ for $Pr<1$ or $\eta_B$ for $Pr>1$.

\begin{figure}
\centerline{\includegraphics[width=1\textwidth]{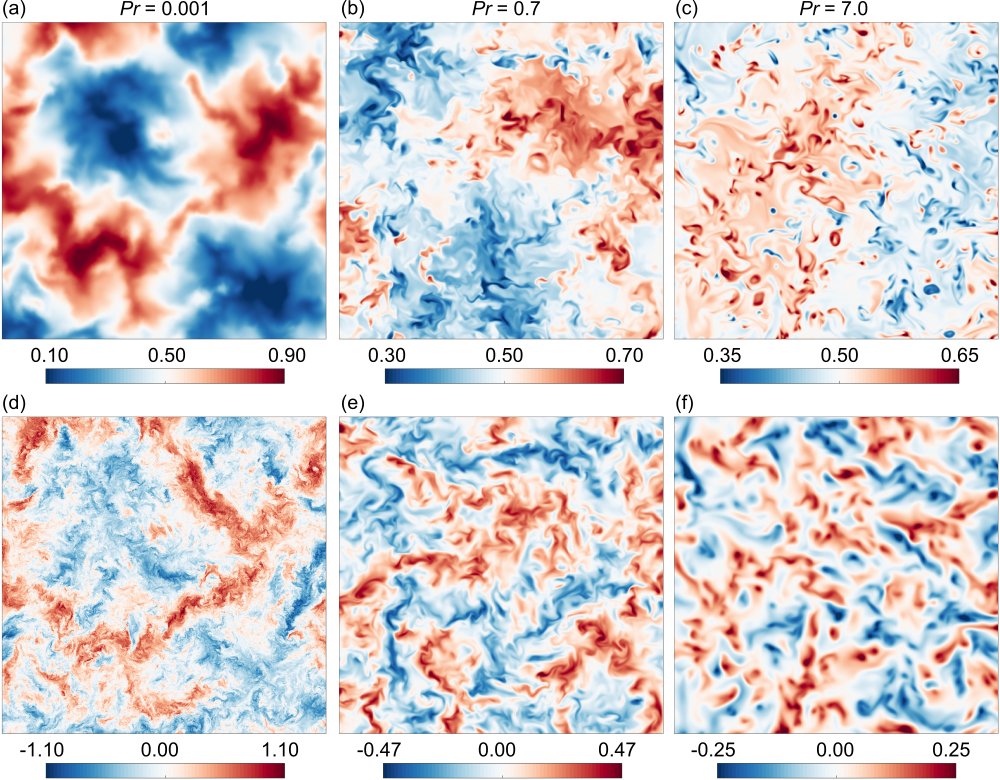}}
\caption{Temperature (top panels) and vertical velocity (bottom panels) in midplane for $Ra = 10^7$. (a,d) are for $Pr = 0.001$, (b,e) for $Pr = 0.7$, and (c,f) for $Pr = 7$. Fields are displayed in a magnified region of dimensions $6.25 H \times 6.25 H$ around the center. Finer temperature contours can be observed for higher $Pr$, panel (c), compared to lower $Pr$, panel (a). The velocity field for lower $Pr$, panel (d), exhibits much finer structures than those in the other two panels, because the Reynolds number decreases as $Pr$ increases.}
\label{fig:T_uz_ra1e7}
\end{figure}

The Corrsin scale is nearly 178 times larger than the Kolmogorov scale for $Pr = 0.001$. This large difference is clear in figures~\ref{fig:T_uz_ke}(c,f), where the temperature and vertical velocity fields are shown in a small cross-section of size $1.56H \times 1.56H$. The figures reveal that the smallest length scale of the thermal structures---the length scale over which the temperature variation is significant---is of the order of $H$, whereas that for velocity structures is much finer. We also find that the dominant structures in the kinetic energy field resemble those in the vertical velocity because of its dominance in the midplane~\citep{Pandey:Nature2018}. This wide range of length scales present in low-$Pr$ convection engenders a broad inertial range in kinetic energy spectrum, which will be discussed in \S~\ref{sec:Ek}.

To see the effects of $Pr$ on flow structures, we show the temperature and the vertical velocity fields for $Pr = 0.001$, $0.7$ and $7$ at $Ra = 10^7$ in figure~\ref{fig:T_uz_ra1e7}. To accentuate small structures, the fields are shown in a quarter (in linear dimension) of the entire cross-section. With increasing $Pr$, increasingly finer thermal structures are generated due to decreasing thermal diffusivity. On the other hand, the velocity variation becomes progressively regular as the viscosity increases (or the Reynolds number decreases) with increasing $Pr$.

\subsection{Heat and momentum transport laws}

\begin{figure}
\centerline{\includegraphics[width=0.9\textwidth]{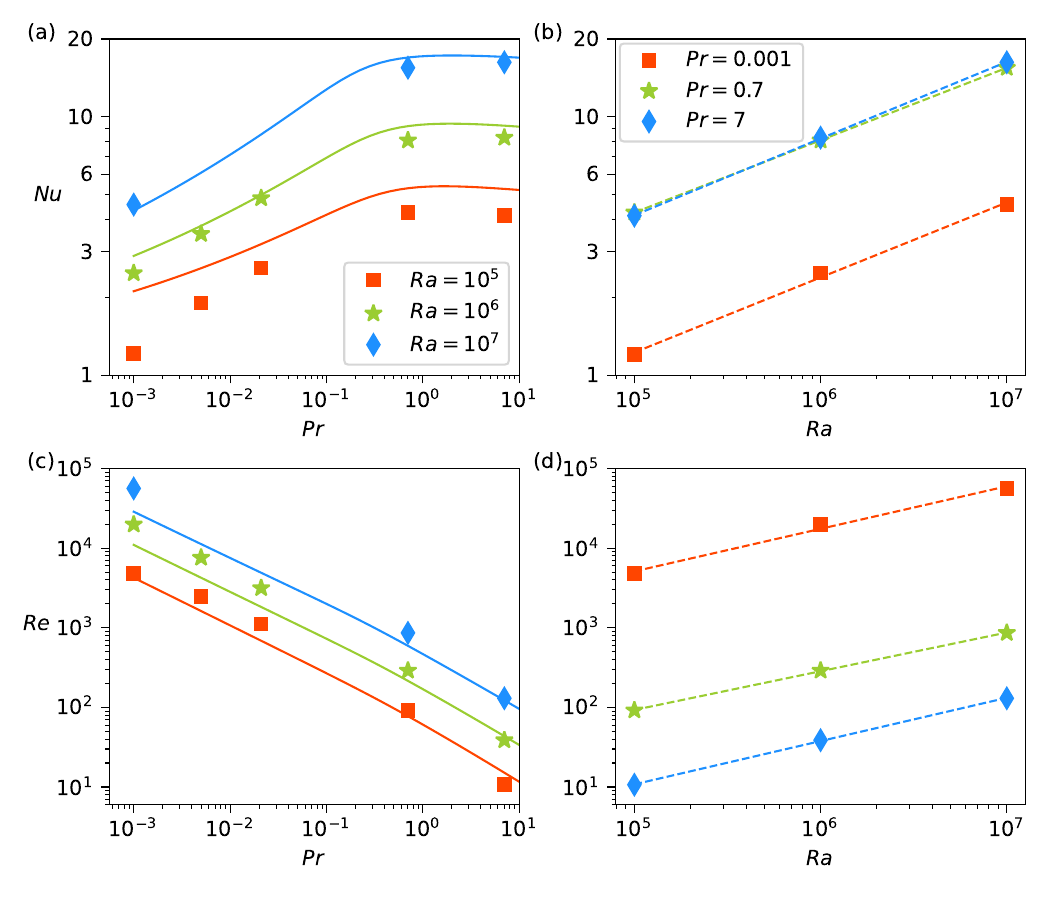}}
\caption{(a) Nusselt number for $\Gamma = 25$ increases with increasing $Pr$ for low Prandtl numbers but does not change much for $Pr \geq 0.7$, consistent with \citet{Pandey:EPL2021}. (b) $Nu$ as a function of $Ra$ increases approximately as $Ra^{0.29}$. (c) The Reynolds number based on the root-mean-square velocity decreases with increasing $Pr$, with power law exponents consistent with others in the literature. (d) $Re$ as a function of $Ra$ increases as $Ra^{0.5}$ for all the three cases. The dashed lines in panels (b) and (d) are best fits, summarized in table~\ref{table:scaling}. Solid lines in panels (a, c) are predictions of $Nu$ and $Re$ from the Grossmann-Lohse theory. Legends in top panels also apply to the corresponding bottom panels.}
\label{fig:Nu_Re}
\end{figure}

Convection at low Prandtl numbers differs from its high-$Pr$ counterpart by reduced heat transport and enhanced momentum transport~\citep{Schumacher:PNAS2015, Scheel:JFM2016, Scheel:PRF2017, Pandey:Nature2018, Zuerner:JFM2019, Zwirner:JFM2020}. Heat transport is quantified by the Nusselt number $Nu$, defined as the ratio of the total heat transport to that by conduction alone. It is computed as 
\begin{equation}
Nu = 1 + \sqrt{RaPr} \, \langle u_z T \rangle_{V,t} \, , \label{eq:Nu}
\end{equation}
where $\langle \cdot \rangle_{V,t}$ denotes again the average over the entire simulation domain and time. We compute $Nu$ for all simulations and plot them, for fixed $Ra$, as a function of $Pr$ in figure~\ref{fig:Nu_Re}(a): $Nu$ increases up to $Pr = 0.7$ but does not change significantly thereafter. A similar trend has also been reported in literature for convection in $\Gamma \approx 1$ domains~\citep{Verzicco:JFM1999, Schmalzl:EPL2004, Poel:JFM2013} and also for $\Gamma = 0.1$ \citep{Pandey:EPL2021}. Figure~\ref{fig:Nu_Re}(a) indicates that the molecular diffusion becomes an increasingly dominant mode of heat transport as $Pr$ decreases. For $Ra = 10^5$ and $10^6$, we do best fits to the data for $Pr \leq 0.7$. The transport laws $Nu(Pr)$ for both Rayleigh numbers are given, including error bars, in table \ref{table:scaling}. In summary, we find that the Nusselt number is consistent with the power law $Nu\sim Pr^{0.19}$ for $Ra = 10^5$ and $Nu\sim Pr^{0.18}$ for $Ra=10^6$. These power law exponents in the extended convection domain are within the range observed in RBC with $\Gamma \lesssim 1$, as shown in \citet{Pandey:EPL2021}, where more discussion of the $Nu-Pr$ scaling exponent can be found. 

Normalized values of global heat transport in RBC increases with increasing thermal driving and the rate of increase depends on the Prandtl number~\citep{Scheel:JFM2016, Scheel:PRF2017}. We plot $Nu$ for $Pr = 0.001, 0.7$, and 7 against $Ra$ in figure~\ref{fig:Nu_Re}(b), which shows that $Nu$ for $Pr = 0.7$ and 7 are nearly similar. There are only three data points, which would not be adequate to establish a new result. Nevertheless, power-law fits to those three points serve to supplement existing results. We obtain approximately $Nu \sim Ra^{0.29}$ (see table \ref{table:scaling}). The exponents for all the three Prandtl numbers are essentially similar and agree with those observed in convection for $\Gamma\sim 1$~\citep{BailonCuba:JFM2010, Stevens:JFM2011, Scheel:JFM2012, Scheel:JFM2014, Scheel:JFM2016}. It is interesting that the exponent for $Pr = 0.001$ is not lower compared to that for $Pr \geq 0.7$. A slightly lower scaling exponent of $0.27\pm 0.01$ was reported from simulations in closed cylinders for $\Gamma=1$ \citep{Scheel:PRF2017} at $Pr=0.021$. Recent experiments in strongly turbulent liquid metal convection by \citet{Schindler:PRL2022}, at nearly the same Prandtl number but for Rayleigh numbers up to $Ra=5\times 10^9$, in a cylinder with $\Gamma=1/2$, reported an even smaller scaling exponent of 0.124. It is possible that the constrained large-scale flow in closed cylinder affects the scaling exponent at low and moderate Rayleigh numbers, see discussion by \citet{Pandey:EPL2021}.

The Reynolds number $Re$ quantifies the momentum transport in RBC. We compute it with $H$ and $u_\mathrm{rms}$ as the relevant length and velocity scales, as
\begin{equation}
Re =  u_\mathrm{rms} \, \sqrt{\frac{Ra}{Pr}} \quad \mbox{where}\quad u_\mathrm{rms} = \sqrt{\langle u_i^2 \rangle_{V,t}}.
\end{equation}
is the root-mean-square (rms) velocity. The Reynolds number as a function of $Pr$ is plotted in figure~\ref{fig:Nu_Re}(c), which reveals that, for fixed $Ra$, the flow loses its effectiveness in transporting momentum as $Pr$ increases~\citep{Kapyla:AA2021}. In thermal convection, the power law exponent of $Re-Pr$ scaling depends on the range of $Pr$; the Reynolds number decreases with increasing $Pr$ even when the flow is dominated by inertia. The detailed power law fits can be found in table \ref{table:scaling}. As a summary, we get $Re \sim Pr^{-0.62}$ and $Re \sim Pr^{-0.65}$ for $Ra = 10^5$ and $Ra = 10^6$, respectively. As for the Nusselt number, the exponents of the $Re-Pr$ scaling agree with those observed for $\Gamma \lesssim 1$ ~\citep{Verzicco:JFM1999, Pandey:EPL2021}. 

The Reynolds number variation with $Ra$, plotted in figure~\ref{fig:Nu_Re}(d), shows that $Re$ is consistently higher for lower Prandtl numbers, manifesting in the enhanced prefactors of the power law fits which are summarized in table \ref{table:scaling}. The best fits yield $Re \sim Ra^{0.53}$, $Re \sim Ra^{0.49}$, and $Re \sim Ra^{0.54}$ for $Pr = 0.001, 0.7$, and 7, respectively. Note that the Reynolds number based on the free-fall velocity scales as $Ra^{0.50}$. Thus, these scaling exponents suggest that the free-fall velocity for a fixed $Pr$ does not depend strongly on $Ra$ (see table~\ref{table:details}). The scaling exponents are in the same range as in several other studies in the past; the exponent does not, however, decrease with decreasing Prandtl number as found by \citet{Scheel:PRF2017}. This might be the result of differences in the aspect ratio, but we reiterate that the present fits use only 3 data points.

Attempts have been made to predict the global transports in RBC as a function of the control parameters \citep{Shraiman:PRA1990, Grossmann:JFM2000, Pandey:POF2016}. \citet{Grossmann:JFM2000} assumed the existence of a large-scale circulation of the order of the size of the convection cell, and proposed a set of coupled equations relating $Nu$ and $Re$ as functions of $Ra$ and $Pr$~\citep{Grossmann:PRL2001}. The equations also include a set of constant coefficients, whose values depend on the aspect ratio of the domain. Using the coefficients provided in \citet{Stevens:JFM2013} for $\Gamma \approx 1$ RBC, we compute $Nu$ as a function of $Pr$ from the Grossmann-Lohse model and show them as solid curves in figure~\ref{fig:Nu_Re}(a). The Nusselt numbers thus estimated are somewhat higher than those computed from the DNS for $Ra = 10^5$. On the low-$Pr$ end, this may be attributed to the fact that the temperature fields for these parameter pairs are dominated by diffusion and barely mixed in the bulk. However, the agreement is better for $Ra \geq 10^6$, which indicates that the heat transport in our extended cell is not much different from that in $\Gamma = 1$ cells. We also plot $Re(Pr)$ from the Grossmann-Lohse model in figure~\ref{fig:Nu_Re}(c), and find that there is fair agreement; see also \citet{Verma:book2018}.

\begin{table}
  \begin{center}
\def~{\hphantom{0}}
  \begin{tabular}{lccc}
$Ra$ & $Pr$ & $Nu$ & $Re$ \\
$10^5$ & $0.001 - 0.7$ & $(4.8 \pm 0.2) Pr^{0.19 \pm 0.02}$ & $(84 \pm 9) Pr^{-0.62 \pm 0.05}$ \\
$10^6$ & $0.001 - 0.7$ & $(9.0 \pm 0.3) Pr^{0.18 \pm 0.01}$ & $(240 \pm 8) Pr^{-0.65 \pm 0.02}$ \\
$10^5 - 10^7$ & 0.001 & $(0.043 \pm 0.004)Ra^{0.29 \pm 0.01}$ & $(10.9 \pm 3.1) Ra^{0.53 \pm 0.05}$ \\
$10^5 - 10^7$ & 0.7 & $(0.17 \pm 0.001)Ra^{0.28 \pm 0.001}$ & $(0.34 \pm 0.02) Ra^{0.49 \pm 0.01}$ \\
$10^5 - 10^7$ & 7 & $(0.14 \pm 0.003)Ra^{0.30 \pm 0.003}$ & $(0.021 \pm 0.001) Ra^{0.54 \pm 0.01}$\\
  \end{tabular}
  \caption{Summary of scaling relations for global heat and momentum transports as functions of $Ra$ and $Pr$. Note that the scaling laws with respect to $Ra$ have been obtained by fits to 3 data points only; it is clear that more definitive results require larger number of data points.}
  \label{table:scaling}
  \end{center}
\end{table}

\section{Vertical profiles across the convection layer}
\label{sec:mean}

\subsection{Temperature and heat flux fields}

\begin{figure}
\centerline{\includegraphics[width=\textwidth]{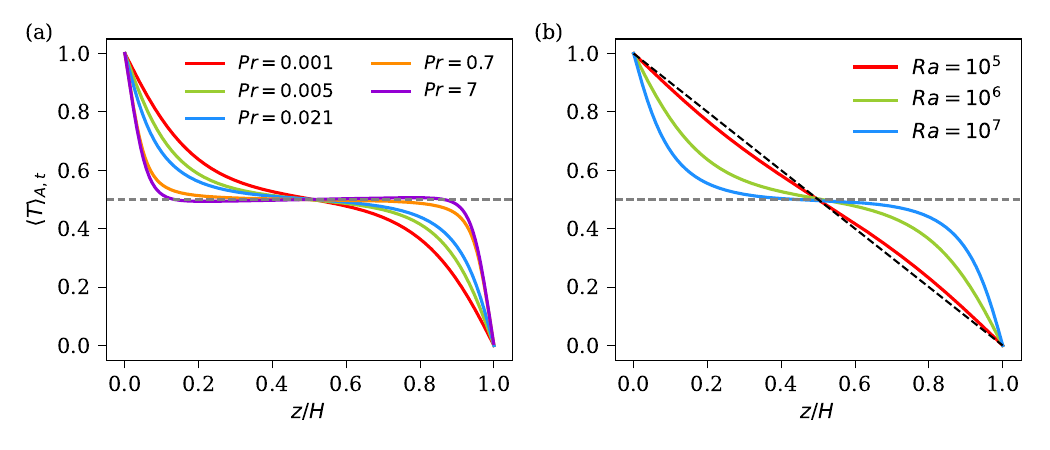}}
\caption{Horizontal and time averages of temperature as a function of the depth for simulations at (a) $Ra = 10^6$ and (b) $Pr = 0.001$. A well-mixed isothermal region away from the walls occurs only for $Pr \geq 0.7$ in (a), whereas a significant temperature gradient in the central region occurs for lower $Pr$. Dashed black line in panel (b) corresponds to the dimensionless conduction temperature profile $T_\mathrm{cond} = 1 - z$.}
\label{fig:T_z1}
\end{figure}

In the conductive equilibrium state, the vertical temperature gradient is a constant; inhomogeneities in the horizontal directions arise in the convective state, leading to a modification of the linear temperature profile. We compute the mean temperature profile $\langle T \rangle_{A,t}(z)$ and plot them in figure~\ref{fig:T_z1}. Here, $\langle \cdot \rangle_{A,t}$ stands for the averaging over the entire horizontal cross-section of $A=25H\times 25H$ at a fixed height $z$ and the full time interval. In a turbulent convective flow, almost the entire temperature drop occurs within the thermal boundary layers (BLs) on the horizontal plates, while the bulk of the flow outside these BLs remains nearly isothermal (and thus well-mixed). Figure~\ref{fig:T_z1}(a) exhibits this feature. However, the slope of the temperature profile in the midplane increases as $Pr$ decreases. We plot the profiles for $Pr = 0.001$ for all the Rayleigh numbers in figure~\ref{fig:T_z1}(b). The profile for $Ra = 10^5$ departs only weakly from the linear conduction profile despite a high Reynolds number of the flow. The temperature gradient in the central plane decreases with increasing $Ra$, and even a Rayleigh number of $10^7$ is not enough to generate a well-mixed temperature field in the bulk region for this very low $Pr$.

\begin{figure}
\centerline{\includegraphics[width=\textwidth]{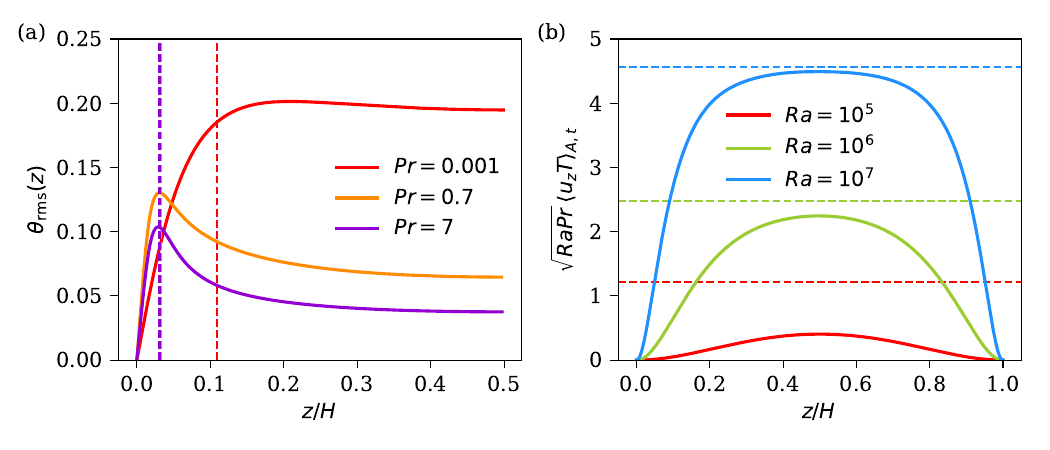}}
\caption{(a) Root-mean-square temperature fluctuation profiles averaged over the top and bottom halves varying with the distance from the plate for $Ra = 10^7$. The peaks in $\theta_\mathrm{rms}(z)$ occur near the thermal BL edge, which are indicated by dashed vertical lines. (b) Vertical profiles of the convective heat flux for $Pr = 0.001$. Dashed horizontal lines indicate the global heat flux ($Nu$) for each case. The convective flux vanishes at the top and bottom plates and is largest in the central plane.}
\label{fig:uzT_th_z}
\end{figure}

In OB convection, the temperature averaged over the entire flow domain is $\Delta T/2$ but fluctuates at each point in the flow. We decompose the temperature field into its mean and fluctuation as 
\begin{equation}
T({\bm x},t) = \langle T \rangle_{A,t}(z) + \theta({\bm x},t) \,.
\end{equation}
Even though the temperature field becomes increasingly diffusive as $Pr$ decreases, the fluctuations increase with decreasing $Pr$; see figure~\ref{fig:uzT_th_z}(a) for $Ra = 10^7$. The depth variation is captured by the planar temperature fluctuation computed as 
\begin{equation}
\theta_\mathrm{rms}(z) = \sqrt{ \langle [T - \langle T \rangle_{A,t}(z)]^2 \rangle_{A,t} } = \sqrt{ \langle T^2(z) \rangle_{A,t} - \langle T(z) \rangle_{A,t}^2}.
\end{equation}
Figure~\ref{fig:uzT_th_z}(a) shows that $\theta_\mathrm{rms}(z)$ vanishes at the plate due to the imposed isothermal boundary condition.
With increasing distance from the bottom plate, however, the strength of fluctuations increases within the thermal BL region. The maxima in $\theta_\mathrm{rms}(z)$ profiles occur near the edge of the thermal BL (computed as $0.5H/Nu$) marked as dashed vertical lines in figure~\ref{fig:uzT_th_z}(a). This suggests that the thermal plumes retain their temperature, while the temperature of the ambient fluid decreases (increases) with increasing distance from the bottom (top) plate. This leads to an increasing contrast between the two components of the flow and is reflected as an increasing $\theta_\mathrm{rms}(z)$ within the BL region~\citep{Pandey:JFM2021}. In the bulk region, however, $\theta_\mathrm{rms}$ decreases with distance from the plate because the plumes do not retain their identity and begin to mix with the bulk fluid. 

The heat transport occurs due to convective as well as diffusive processes, with their ratio varying with depth. To get the total heat flux in a horizontal plane, we average the temperature equation~(\ref{eq:T}) in horizontal directions and in time, which leads to
\begin{equation}
Nu(z) = \sqrt{Ra Pr} \langle u_z T \rangle_{A,t} - \frac{\partial \langle T \rangle_{A,t}}{\partial z}=\mbox{const}\,. \label{eq:Nuz}
\end{equation}
It is clear from the temperature profiles in figure~\ref{fig:T_z1} that the diffusive contribution $- \partial \langle T \rangle_{A,t}/\partial z$ should be small in the well-mixed bulk region---increasing towards the plates and becoming largest at the plates. The variation of the convective heat flux $\sqrt{Ra Pr} \langle u_z T \rangle_{A,t}$ with depth in figure~\ref{fig:uzT_th_z}(b) for $Pr = 0.001$ confirms this expectation. The magnitudes of the globally-averaged heat flux are indicated as dashed horizontal lines in figure~\ref{fig:uzT_th_z}(b), showing that the diffusive flux (the distance between the solid curves and the corresponding dashed horizontal lines) is not negligible even in the central region for $Ra \leq 10^6$. The diffusive component dominates the total heat flux in the central region for $Ra = 10^5$, which is consistent with the highly inefficient convective heat transport; see table~\ref{table:details}. However, for $Ra = 10^7$, the diffusive contribution diminishes in the central plane. Thus, as $Pr$ becomes smaller, one requires increasing $Ra$ before turbulent processes become important.

\subsection{Thermal and kinetic energy dissipation rates}

While the mean temperature $\langle T\rangle_{A,t}(z)$ varies sharply near the horizontal plates and weakly in the central region, the vertical mean profile of the thermal dissipation rate field, which is the rate of loss of thermal variance that is  computed pointwise by
\begin{equation}
\varepsilon_T({\bm x},t) = \kappa \left[ \left( \frac{\partial T}{\partial x} \right)^2 + \left( \frac{\partial T}{\partial y} \right)^2 + \left( \frac{\partial T}{\partial z} \right)^2 \right]\,,
\end{equation}
is higher in the vicinity of the horizontal plates and decreases towards the centre~\citep{Scheel:JFM2016}.
\begin{figure}
\centerline{\includegraphics[width=\textwidth]{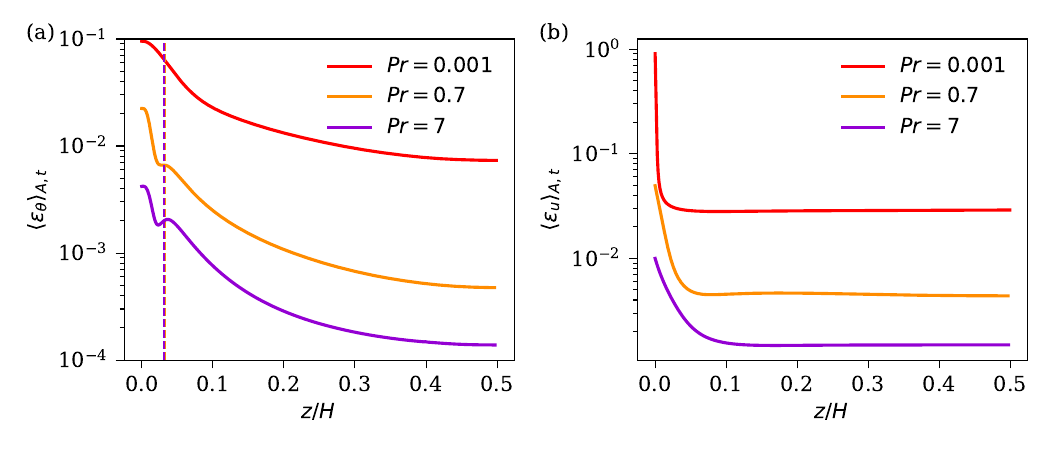}}
\caption{Variation of the horizontally averaged (a) thermal and (b) kinetic energy dissipation rates in the vertical direction for $Ra = 10^7$. The profiles are further averaged over the top and bottom halves of the domain to improve the statistics. The dissipation profiles are largest at the plates and decrease towards the central plane; however, the $\varepsilon_u(z)$ is nearly uniform in the bulk region. The (indistinguishable) dashed vertical lines in panel (a) indicate the edges of the thermal BLs for $Pr = 0.7$ and $Pr = 7$.}
\label{fig:eps_z}
\end{figure}
We also compute the thermal dissipation rate field defined as
\begin{equation}
\varepsilon_\theta({\bm x},t) = \kappa \left[ \left( \frac{\partial \theta}{\partial x} \right)^2 + \left( \frac{\partial \theta}{\partial y} \right)^2 + \left( \frac{\partial \theta}{\partial z} \right)^2 \right]
\end{equation}
to quantify the spatial variation of the temperature fluctuations. The mean profile of the thermal dissipation rate $\langle \varepsilon_\theta \rangle_{A,t}(z)$ is plotted in figure~\ref{fig:eps_z}(a) for $Ra = 10^7$. Note that the vertical mean profiles of $\varepsilon_T$ and $\varepsilon_\theta$ are related by
\begin{equation}
\langle \varepsilon_T \rangle_{A,t}(z) = \langle \varepsilon_\theta \rangle_{A,t}(z) + \varepsilon_{\langle T \rangle}(z),
\end{equation}
where $\varepsilon_{\langle T \rangle} =  \kappa ({\rm d} \langle T \rangle_{A,t}/{\rm d} z)^2$ is the dissipation rate corresponding to the mean temperature profile~\citep{Emran:JFM2008}. In convective flows with well-developed thermal BLs, $\varepsilon_{\langle T \rangle}$ contributes primarily to the boundary layers and negligibly in the bulk. This rapid decrease of $\varepsilon_{\langle T \rangle}$ outside the thermal BL region shows a shallow kink in the profiles of $\langle\varepsilon_\theta\rangle_{A,t}(z)$. In figure~\ref{fig:eps_z}(a), we indicate the thermal BL thicknesses for $Pr = 0.7$ and $Pr = 7$ as dashed vertical lines, and note that the kinks are observed near the edge of the thermal BL. The kink does not appear for $Pr = 0.001$ due to the absence of well-developed thermal BLs. 

We find that $\langle \varepsilon_\theta\rangle_{A,t}(z)$ increases with decreasing $Pr$. This is because the volume-averaged thermal dissipation rate is related to the global heat transport \citep{Shraiman:PRA1990} as
\begin{equation}
\langle \varepsilon_T \rangle_{V,t} = \frac{Nu}{\sqrt{Ra Pr}} \, .
\end{equation}
As we observe $Nu \sim Pr^{0.2}$, this leads to $\langle \varepsilon_T \rangle_{V,t} \sim Pr^{-0.3}$ for a fixed $Ra$. Thus, the decrease of the thermal dissipation rate with increasing $Pr$ is consistent with the $Pr$-dependence of the Nusselt number. 

We now plot in figure~\ref{fig:eps_z}(b) the profiles of the viscous dissipation rate defined in~(\ref{eq:diss_u}). Similar to $\langle\varepsilon_\theta\rangle_{A,t}(z)$, the largest values of $\langle\varepsilon_u\rangle_{A,t}(z)$ are found near the horizontal plate owing to the strongly varying velocity field in the vicinity of the plates. Further, the variation of the profiles $\langle\varepsilon_u\rangle_{A,t}(z)$ in the bulk region is almost negligible compared to that in the viscous BL region near the plates. Figure~\ref{fig:eps_z}(b) shows that $\langle\varepsilon_u\rangle_{A,t}(z)$ increases with decreasing $Pr$ for all $z$. Note that the globally-averaged viscous dissipation rate is related to the Nusselt number as 
\begin{equation}
\langle \varepsilon_u \rangle_{V,t} = \frac{Nu-1}{\sqrt{RaPr}} \, ,
\end{equation}
and therefore, $\langle \varepsilon_u \rangle_{V,t}$ should decrease with increasing $Pr$. 

\section{Characterization of the turbulence in the bulk}
\label{sec:bulk}

\subsection{Isotropy in the midplane}
\citet{Vorobev:POF2005} used the ratios 
\begin{equation}
    G_{ij}=\frac{\langle(\partial u_i/\partial z)^2\rangle (1+\delta_{iz})}{\langle(\partial u_i/\partial x_j)^2\rangle (1+\delta_{ij})} \quad\mbox{with}\quad i,j=x,y,z\,
    \label{Gij}
\end{equation}
to determine the degree of anisotropy on the level of second-order derivative moments. Flows with no variation in the vertical direction $z$ yield $G_{ij}\to 0$ (and are thus anisotropic), while $G_{ij} = 1$ for perfectly isotropic flows. The coefficient $G_{11}$, relating the in-plane derivative to a transverse derivative with respect to the vertical direction is summarized in three horizontal planes in table~\ref{table:Gij}. $G_{11}$ remains nearly unity in the bulk region for $0.1 \leq z/H \leq 0.9$, but significant departures are found near the horizontal plates. Similar amplitudes follow for other combinations; see also \citet{Nath:PRF2016}.  We thus conclude that a plausible case exists for exploring similarities with Kolmogorov turbulence in the bulk region; see~\citet{Mishra:PRE2010} and~\citet{Verma:NJP2017}.
\begin{table}
  \begin{center}
\def~{\hphantom{0}}
  \begin{tabular}{lccccc}
$z/H$ & $(Ra,Pr)$ & $(Ra,Pr)$ & $(Ra,Pr)$ & $(Ra,Pr)$ & $(Ra,Pr)$ \\
 & $10^5, 0.001$ & $10^6, 0.001$ & $10^7, 0.001$ & $10^6, 0.005$ & $10^6, 0.021$ \\
0.5 & 0.971 & 0.980 & 0.957 & 0.953 & 0.963 \\
0.1 & 1.132 & 1.008 & 1.018 & 1.056 & 1.173 \\
0.01 & 11.35 & 3.225 & 1.657 & 7.108 & 12.51 \\
  \end{tabular}
  \caption{The anisotropy coefficient $G_{11}$ in three horizontal planes; for definition see \eqref{Gij}. $G_{11}$ as well as the other coefficients $G_{ij}$ remain close to unity in the bulk region between $z = 0.1H$ and $z = 0.9H$, but depart significantly as the horizontal plate is approached and the shear effects dominate.}
  \label{table:Gij}
  \end{center}
\end{table}

\subsection{Kinetic energy spectra}
\label{sec:Ek}

In three-dimensional turbulent flows, the kinetic energy injected at large length scales cascades towards smaller scales and eventually gets dissipated at the smallest scales by viscous action.\footnote{We shall not consider the connection to the Onsager conjecture that it may be related to singularities in weak solutions of the Euler equations.} In the inertial range---the range of length scales far from both the injection as well as dissipation scales---the kinetic energy spectrum $E(k)$, whose integral over all wavenumbers $k$ yields the kinetic energy, follows the standard Kolmogorov scaling
\begin{equation}
 E(k) = K_\mathrm{K} \varepsilon_u^{2/3} k^{-5/3},	\label{eq:Ek_Kol}
\end{equation}
where $K_\mathrm{K}$ is the Kolmogorov constant and $\varepsilon_u$ denotes the volume- and time-averaged kinetic energy dissipation rate. 

For our purposes, it would be useful to study the behaviour of two-dimensional (2D) energy spectrum in a horizontal plane. The 2D Fourier transform of a field $f(x,y,z_0)$ in a horizontal plane at $z = z_0$ is defined as 
\begin{equation}
f(x,y,z_0) = \int_{-\infty}^{\infty} \int_{-\infty}^{\infty}  \hat{F}(k_x,k_y) e^{-i (k_x x + k_y y)} dk_x dk_y,
\end{equation}
where $\hat{F}(k_x,k_y)$ is the Fourier mode corresponding to the wavevector ${\bm k} \equiv (k_x,k_y)$. Thus, the Fourier modes of the velocity field in midplane ${\bm u}(x,y,z=H/2) \equiv {\bm U}(x,y)$ are denoted as $\hat{\bm U}({\bm k}) \equiv [\hat{U}_x({\bm k}), \hat{U}_y({\bm k}), \hat{U}_z({\bm k})]$. The kinetic energy in a horizontal plane is equal to the sum of the energies of each Fourier mode, i.e.,
\begin{align}
\frac{1}{2} \langle {\bm U}^2 \rangle_{A,t} &= \int_{-\infty}^{\infty} \int_{-\infty}^{\infty} \frac{1}{2} |\hat{\bm U}(k_x,k_y)|^2 dk_x dk_y \nonumber\\
&= \int_{0}^{\infty} \left[ \int_{0}^{2\pi} \frac{1}{2}  |\hat{\bm U}(k,\phi_k)|^2 d\phi_k \right] k dk,
\end{align}
with
\begin{equation}
k = \sqrt{k_x^2+k_y^2} \quad\mbox{and}\quad \phi_k = \arctan(k_y,k_x)\,.
\end{equation}
Using the horizontal isotropy of the fields, the expression in the square brackets could be readily integrated to yield $\pi |\hat{\bm U}(k)|^2$, where $|\hat{\bm U}(k)|^2/2$ is the average kinetic energy of all the Fourier modes lying in an annular region between radii $k$ and $k+dk$. Thus, the average planar kinetic energy becomes
\begin{equation}
\frac{1}{2} \langle {\bm U}^2 \rangle_{A,t} = \int_{0}^{\infty} \pi k |\hat{\bm U}(k)|^2 dk = \int_{0}^{\infty} E(k) dk,
\label{spec}
\end{equation}
where $E(k) = \pi k |\hat{\bm U}(k)|^2$ is the one-dimensional (1D) kinetic energy spectrum in a horizontal plane~\citep{Peltier:JAS1996}.

\begin{figure}
\centerline{\includegraphics[width=\textwidth]{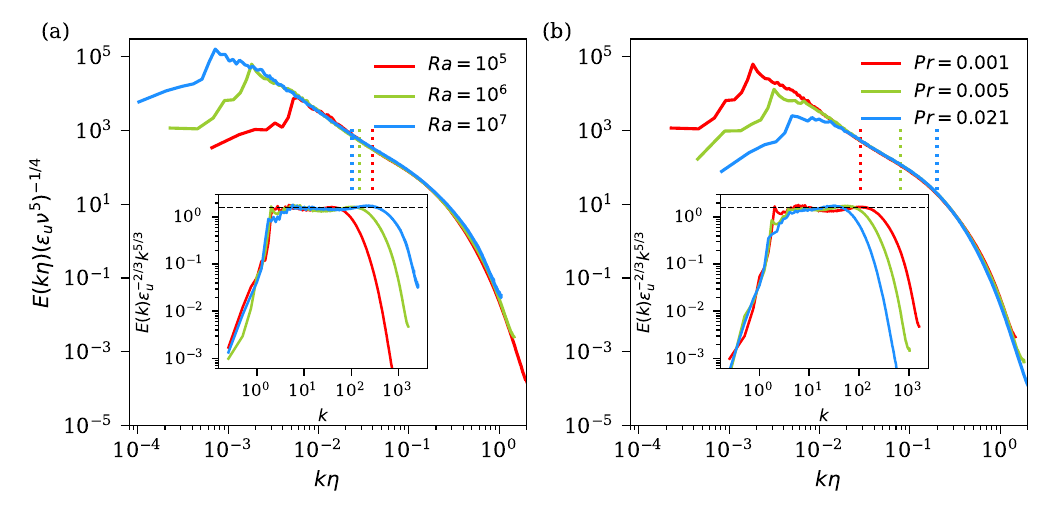}}
\caption{Normalized kinetic energy spectra $E(k\eta)(\varepsilon_u \nu^5)^{-1/4}$ in the midplane as a function of the normalized wavenumber $k\eta$ for $Pr = 0.001$ (a) and $Ra = 10^6$ (b) collapse well in the inertial as well as viscous ranges. Here, $\varepsilon_u = \langle \varepsilon_u \rangle_{V,t}$. The short dotted vertical lines indicate the normalized wavenumber $k_T \eta$, where $k_T = 2\pi/\delta_T$ is the wavenumber corresponding to the mean thermal boundary layer thickness $\delta_T=H/(2 Nu)$. Insets show that the corresponding spectra normalized with $\varepsilon_u^{2/3} k^{-5/3}$ exhibit plateau in the inertial range, which becomes wider with increasing $Ra$ or decreasing $Pr$. The dashed horizontal line in the insets yields the Kolmogorov constant $K_\mathrm{K} \approx 1.6$ in the planar energy spectra, consistent with the value for isotropic turbulence \citep{Sreenivasan:POF1995}.}
\label{fig:Ek}
\end{figure}
We compute the energy spectrum in the midplane of the low-$Pr$ flows for each instantaneous snapshot and then average the instantaneous spectra over all the available snapshots to obtain the mean kinetic energy spectrum. The energy spectra for flows with small-scale universality at different Reynolds numbers should collapse at sufficiently high Reynolds number if they are plotted against $k \eta$. Equation~(\ref{eq:Ek_Kol}) in terms of the normalized wavenumber $k \eta$ reads then~\citep{moninandyaglom} as
\begin{equation}
E(k\eta) = K_\mathrm{K} (\varepsilon_u \nu^5)^{1/4} (k\eta)^{-5/3}\,.
\end{equation}
We now plot the normalized energy spectra $E(k\eta) (\varepsilon_u \nu^5)^{-1/4}$ as a function of $k\eta$ in figure~\ref{fig:Ek}. The spectra for $Pr = 0.001$ are shown in figure~\ref{fig:Ek}(a) for all three Rayleigh numbers, whereas the spectra for a fixed $Ra = 10^6$ and $Pr = 0.001, 0.005$, and 0.021 are displayed in figure~\ref{fig:Ek}(b). The collapse is excellent beyond the wavenumber corresponding to the maximum of $E(k\eta)$. We show the same spectra in the normalized form $E(k) \varepsilon_u^{-2/3} k^{5/3}$ in the insets of figure~\ref{fig:Ek}, which confirm that they collapse on each other for all the low-$Pr$ flows in low to moderately large wavenumber range. The plateau in the dimensionless wavenumber range $k \in [k_0, k_1]$ increases with increasing $Ra$. The inset of figure~\ref{fig:Ek}(a) shows that the inertial range $[k_0, k_1]$ can be estimated to be $[2, 60]$, $[4, 150]$, and $[6, 300]$ for $Ra = 10^5, 10^6, 10^7$ and $Pr = 0.001$, respectively. The inertial range increases with increasing $Ra$ and decreasing $Pr$, as expected from the increased Reynolds numbers. 

The energy spectra normalized in the same way for $Pr = 0.001, 0.005$, and $0.021$ for a fixed $Ra = 10^6$ are shown in the inset of figure~\ref{fig:Ek}(b). Again, the normalized spectra collapse quite well. A plateau can be detected for the two lower $Pr$ with the inertial range corresponding to $[4, 150]$ and $[4, 90]$ for $Pr = 0.001$ and $Pr = 0.005$, respectively. The plateau for all cases corresponds to $K_\mathrm{K} \approx 1.6$, consistent with the experimental and numerical value in isotropic turbulence \citep{Sreenivasan:POF1995,Yeung:PRE1997,Ishihara:ARFM2009}. 
\begin{figure}
\centerline{\includegraphics[width=0.6\textwidth]{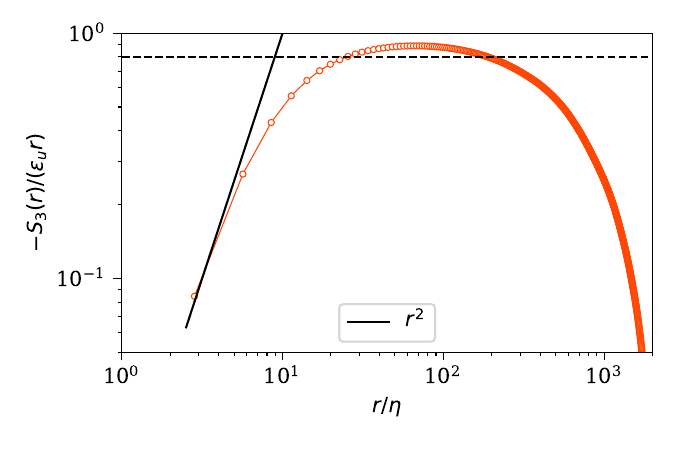}}
\caption{Third-order longitudinal structure function (averaged over the $x$- and $y$-directions) in the midplane for $Pr = 0.001, Ra = 10^7$ varies as $r^2$ in the viscous range, whereas it remains nearly a constant in the inertial range. Dashed horizontal line indicates the constant 4/5 appearing in equation~(\ref{eq:third}). Here, $\varepsilon_u = \langle \varepsilon_u \rangle_{A,t}(z=H/2)$.}
\label{fig:S3}
\end{figure}

\subsection{Third-order structure function}
To further explore whether the velocity fluctuations in the bulk of low-$Pr$ convection are close to Kolmogorov turbulence, we compute the third-order longitudinal structure function defined as 
\begin{equation}
S_3(r) = \langle (\delta_L u(r))^3 \rangle,
\end{equation}
where $\langle \cdot \rangle$ denotes an appropriate averaging, and the longitudinal velocity increment is
\begin{equation}
\delta_L u(r) = [{\bm u}({\bm x} + {\bm r}) - {\bm u}({\bm x})] \cdot \hat{\bm r}\,,
\end{equation}
with $\hat{\bm r}={\bm r}/r$. \citet{Kolmogorov:DANS1941c} showed that $S_3(r)$ in high-$Re$ homogeneous and isotropic turbulent flow is a universal function of the separation $r$ and varies as
\begin{equation}
S_3(r) = -\frac{4}{5} \varepsilon_u r	\label{eq:third}
\end{equation}
in the inertial range; here and for the remainder of the work $\varepsilon_u:=\langle \varepsilon_u \rangle_{A,t}(z=1/2)$. The longitudinal structure functions in the $x$- and $y$-directions in midplane for our highest-$Re$ flow show that they are nearly the same for small and moderate increments, so we average in the two directions. We show in figure~\ref{fig:S3} the averaged third-order structure function in the normalized form $-S_3(r)/(\varepsilon_u r)$ as a function of $r/\eta$. The figure shows that the compensated structure function tends to scale with the analytical form of $r^2$ in the beginning of the viscous range at $r/\eta\sim 1$. It also shows that the normalized structure function exhibits a plateau for an intermediate range of length scales, which implies that $S_3(r)$ indeed approximately varies as $r$ in the inertial range but the numerical value is slightly larger than $4/5$ \citep{Kolmogorov:DANS1941c}. One possible reason for this departure is the remnant buoyancy contribution~\citep{Yakhot:PRL1992}.

\begin{figure}
\centerline{\includegraphics[width=0.7\textwidth]{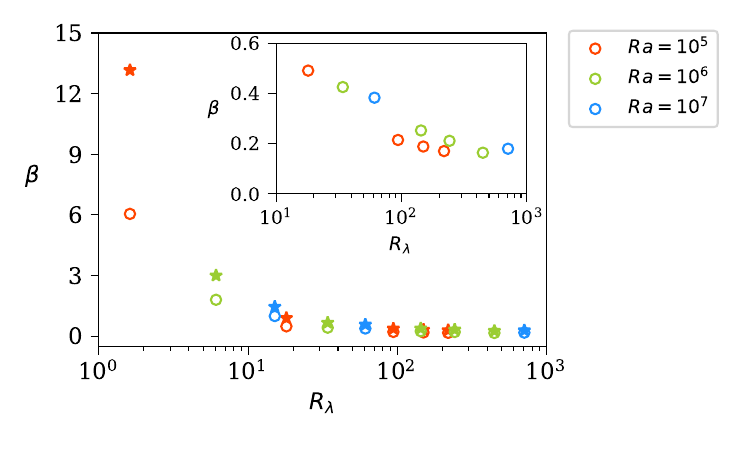}}
\caption{Test of the dissipative anomaly in the bulk of the convection layer using the rescaled mean kinetic energy dissipation rate $\beta$, as given by \eqref{beta}, versus the Taylor microscale Reynolds number. Asterisks use the integral scale ${\cal L}=\ell$ in $\eqref{beta}$, and open circles use ${\cal L}=H$. The inset expands the scale for the case ${\cal L} = H$. Symbols from left to right in each color correspond to decreasing $Pr$. The Taylor microscale Reynolds number is given by \eqref{Rlambda}.}
\label{diss}
\end{figure}

\subsection{Dissipative anomaly}

The zeroth law of turbulence (or ``dissipative anomaly") states that the mean kinetic energy dissipation rate when scaled by a large-scale velocity such as the root-mean-square and the large length scale becomes a constant for sufficiently high Reynolds numbers~\citep{Eyink:PD1994, Frisch:book, Vincent:PRE2021}. Figure \ref{diss} displays this rescaled energy dissipation rate
\begin{equation}
    \beta=\frac{\varepsilon_u {\cal L}}{u_{\rm rms}^3}\,,
    \label{beta}
\end{equation}
where ${\cal L}$ is either the height $H$ of the convection layer or the integral scale $\ell$ calculated from the energy spectrum \eqref{spec} in \S~\ref{sec:Ek} by
\begin{equation}
    \ell=2\pi\,\dfrac{\int_0^{\infty} k^{-1} E(k) dk}{\int_0^{\infty} E(k) dk}\,.
\end{equation}
Note that $u_{\rm rms}$ is also taken with respect to the midplane. The figure summarizes both versions of $\beta$ versus the Taylor microscale Reynolds number 
\begin{equation}
R_{\lambda}= \left( \frac{25 Ra}{9\varepsilon_u^2 Pr} \right)^{1/4} \, u_{\rm rms}^2
\label{Rlambda}
\end{equation}
in the bulk region of the flow. The data for different Rayleigh and Prandtl numbers collapse nicely on a curve that saturates at an approximate value of 0.2 (see the inset of the figure). While this is smaller than 0.45 found by \citet{Sreenivasan:POF1998} for isotropic turbulence, the result implies that the strongly disparate viscous and thermal boundary layer widths (and thus the plume stem widths) do not matter for the driving of the turbulence cascade in the bulk of the flow. However, it also highlights the intrinsic difference in dissipation between convection and isotropic turbulence.

\section{Characteristic lengths and times of turbulent superstructures}
\label{sec:TSS}
The characteristic length scale of the dominant energy-containing structures is of the order of the size of the system when $\Gamma \approx 1$, e.g., a large-scale circulation covering the entire domain ~\citep{Schumacher:PRF2016, Zuerner:JFM2019, Zwirner:JFM2020}. However, for $\Gamma\gg 1$, mean circulation rolls with diameters  larger than $H$ are observed~\citep{Emran:JFM2015}; the resulting large-scale patterns of these rolls are termed  turbulent superstructures of convection~\citep{Pandey:Nature2018, Stevens:PRF2018, Fonda:PNAS2019, Green:JFM2020} as we stated already in \S~\ref{sec:intro}. Although the superstructures extend all the way from the bottom to the top plate~\citep{Pandey:Nature2018}, they are conspicuous when the vertical velocity $u_z$, temperature $T$, or the vertical heat flux $u_zT$ fields are visualized in the midplane~\citep{Fonda:PNAS2019, Green:JFM2020}. For instance, figure~\ref{fig:T_uz_ke} shows turbulent superstructures for low-$Pr$ convection in the form of hot upflows and cold downflows.

The characteristic length $\lambda$ of the superstructures is the typical distance between two consecutive upwelling or downwelling regions and can be estimated using one-dimensional (1D) spectra of the thermal variance, kinetic energy, and convective heat flux in a horizontal plane~\citep{Hartlep:PRL2003, Pandey:Nature2018, Stevens:PRF2018, Green:JFM2020, Krug:JFM2020}. It can also be estimated by computing the two-point auto-correlation function of $u_z$ or $T$ in a horizontal plane and identifying the location of the first minimum, corresponding to $\lambda/2$~\citep{Pandey:Nature2018}. 
\begin{figure}
\centerline{\includegraphics[width=1.0\textwidth]{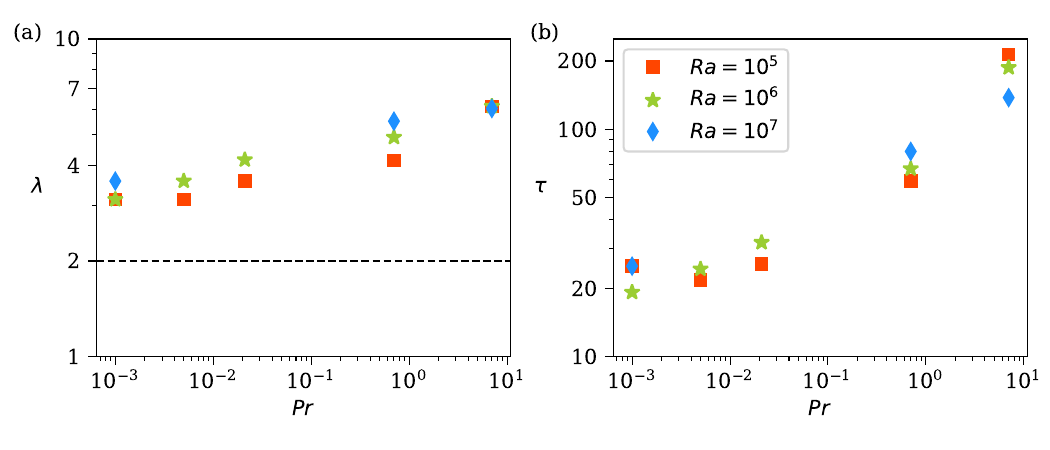}}
\caption{Characteristic spatial (a) and temporal (b) scales of the turbulent superstructures as a function of $Pr$. Both temporal and spatial scales decrease with decreasing Prandtl number, but seem to level off for $Pr \leq 0.005$.}
\label{fig:lambda_tau}
\end{figure}

We compute the 1D power spectra of the vertical velocity, temperature, and convective heat flux, all averaged with respect to the azimuthal angle. They are given by
\begin{align}
S_U(k) & =  \frac{1}{2\pi} \int_0^{2\pi} |\hat{U}_z(k,\phi_k)|^2 d \phi_k, \label{su}\, \\
S_\Theta(k) & =  \frac{1}{2\pi} \int_0^{2\pi} |\hat{\Theta}(k,\phi_k)|^2 d \phi_k, \label{st}\, \\
S_{U\Theta}(k) & =  \frac{1}{2\pi} \int_0^{2\pi} \Re[\hat{U}_z^*(k,\phi_k) \hat{\Theta}(k,\phi_k)] d \phi_k, \label{sut}\,
\end{align}
where $\hat{U}_z(k,\phi_k)$ and $\hat{\Theta}(k,\phi_k)$ are, respectively, the 2D Fourier transforms of $u_z(x,y,H/2)$ and $\theta(x,y,H/2)$. Note that the spectrum $S_U(k)$ differs from the energy spectrum $E(k)$ defined in \S~\ref{sec:Ek}. We find that the three spectra exhibit a peak at nearly the same wavenumber $k^\omega_\mathrm{max}$ corresponding to the maximum of $S_\omega(k)$ (see figure~\ref{fig:E_k_pr0.005}), yielding the characteristic spatial scale $\lambda_\omega = 2\pi/k^\omega_\mathrm{max}$ of the superstructures, with $\omega = \lbrace U, \Theta, U\Theta \rbrace$. 

\begin{figure}
\centerline{\includegraphics[width=0.9\textwidth]{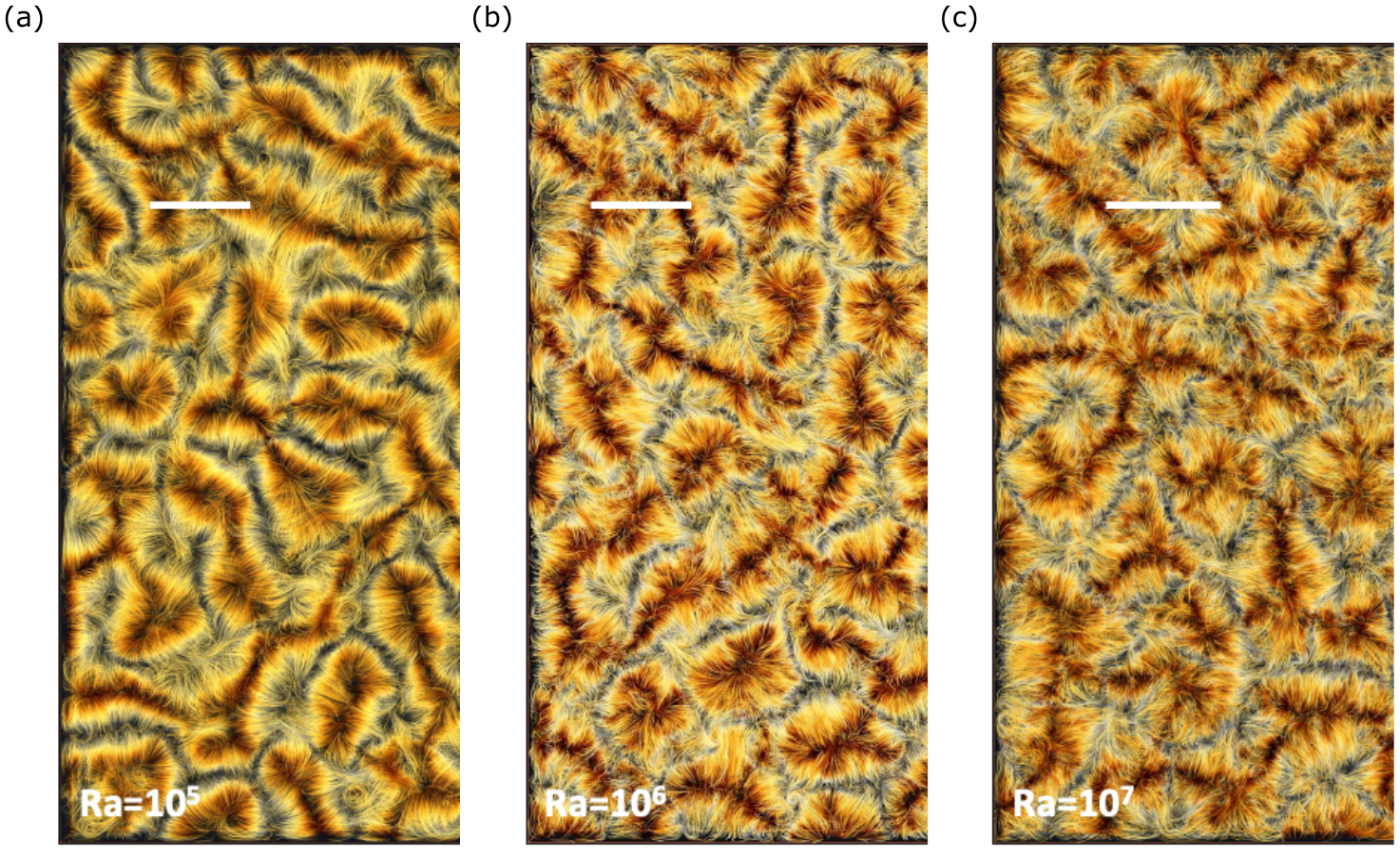}}
\caption{Plots of streamlines for snapshots at (a) $Ra=10^5$, (b) $10^6$, and (c) $10^7$, all at $Pr=0.001$. The view is from the top. Only one half of the cross-section is shown. The corresponding values of $\lambda$ are indicated by a horizontal white bar in each panel.}
\label{fig:stream}
\end{figure}

\citet{Pandey:Nature2018, Pandey:APJ2021} found that the characteristic scales $\lambda_U$ and $\lambda_\Theta$ do not always agree with each other, the contrast being usually larger at moderate and high Prandtl numbers. Here, we extract the spatial scale of superstructures from the power spectrum of the convective heat flux (see also \citet{Hartlep:PRL2003} and \citet{Krug:JFM2020}). It was observed in~\citet{Pandey:Nature2018} that $\lambda$ is a function of $Pr$ and the maximum of $\lambda(Pr)$ was found at $Pr \approx 7$ for $Ra = 10^5$. In the present simulations (in which the $Pr$ range has been extended down to $0.001$ and $Ra$ by two orders of magnitude), the corresponding values of $\lambda$ as a function of $Pr$ are plotted in figure~\ref{fig:lambda_tau}(a) for all simulations of table \ref{table:details}. We find that $\lambda$ decreases with decreasing $Pr$ but they seem to level off at $\lambda \simeq 3H$ for the lowest Prandtl numbers of 0.005 and 0.001; see also figure \ref{fig:stream} where the magnitude of $\lambda$ is displayed in the streamline plot. It remains to be studied as to how this scale arises starting from the onset of RBC where $\lambda = 2.02 H$ independently of $Pr$ \citep{Chandrasekhar:book}, as indicated in figure~\ref{fig:lambda_tau}(a) as a dashed horizontal line. We also refer to Appendix C for further details.

The characteristic temporal scale ($\tau$) of the superstructures is long compared to the time scale of the turbulent fluctuations. Thus, time-averaging the velocity or temperature fields over a time interval $\tau$ yields coarse-grained fields, which are nearly devoid of the small-scale fluctuations. The coarse-grained large-scale structures evolve on the time scales which are of the order of $\tau$~\citep{Pandey:Nature2018, Fonda:PNAS2019}, related to the time for a fluid parcel to complete a circulation. It is computed as 
\begin{equation}
\tau = 3 \frac{\pi(\lambda/4 + H/2)}{u_\mathrm{rms}}, \label{eq:tau}
\end{equation}
where the quantity in the numerator is the circumference of superstructure rolls with elliptical cross-section~\citep{Pandey:Nature2018}. The factor of three in the above expression arises from the fact that the circulation time in an extended convection flow is not fixed but exhibits a broad distribution with stretched-exponential tails in the Lagrangian frame of reference along massless tracer trajectories~\citep{Schneide:PRF2018,Vieweg:PRF2021}. The scale computed using expression~(\ref{eq:tau}) is plotted in figure~\ref{fig:lambda_tau}(b) as a function of $Pr$. We find that $\tau$ increases with $Pr$, consistent with the fact that the Reynolds number, and thus the characteristic velocity $u_\mathrm{rms}$, decreases with increasing $Pr$, requiring a longer time for fluid parcels to complete a circulation. Note that $\lambda$ in equation~(\ref{eq:tau}) also increases with increasing $Pr$, but the increase is not as significant as the decrease in $u_\mathrm{rms}$.

\section{Conclusions and outlook}
\label{sec:concl}

Our focus here has been the Rayleigh-B\'{e}nard convection in a horizontally extended layer for molecular Prandtl numbers as small as $Pr=10^{-3}$, which go beyond those accessible in controlled laboratory experiments and approach astrophysical conditions. We extended the parameter space of previous works by \citet{Pandey:Nature2018} and \citet{Fonda:PNAS2019} by direct numerical simulations, both towards lower $Pr$ and higher $Ra$, and thus determined more conclusively various parameter dependencies such as global heat and momentum transports, temperature fluctuations, as well as the kinetic and thermal dissipation rates. Among others, these results provide a test for existing predictions by the theory of \citet{Grossmann:PRL2001} and \citet{Stevens:JFM2013}. Comparisons show that the predictions for the global heat and momentum transports as a function of the Prandtl number at fixed Rayleigh number are in fair agreement. 

We also found that the Nusselt number decreases as $Nu \sim Pr^{0.19}$, whereas the Reynolds number increases approximately as $Re \sim Pr^{-0.6}$ when the Prandtl number decreases from $Pr = 0.7$ to 0.001 as detailed in table \ref{table:scaling}. The dimensionless mean thermal and kinetic energy dissipation rates also decrease with increasing $Pr$ and their scaling behaviors are consistent with that of the global heat transport. We studied the depth dependence of these quantities, and found that, due to a high diffusive temperature field in convection at very low $Pr$, the bulk fluid is not mixed well and a significant vertical temperature gradient occurs in the bulk region, even for the highest accessible Rayleigh number.

The highly inertial fluid turbulence in the bulk of low-$Pr$ convection layer was studied by examining the kinetic energy spectra, the 4/5-ths law, local isotropy, and a test of the dissipative anomaly. The results suggest that the fluid turbulence in the bulk for the lowest Prandtl number is close to the classical Kolmogorov turbulence. This implies that the temperature field behaves as a passive and highly diffusive scalar stirred by a highly turbulent flow; the impact of thermal plumes, which are the unstable fragments of the thick thermal boundary layers, can be considered an efficient large-scale forcing for turbulence, dominantly in the lower wavenumber part of the inertial range ($k_T$ in figure \ref{fig:Ek}). Some differences do remain. In particular, the asymptotic value of the rescaled mean kinetic energy dissipation rate falls below that in isotropic turbulence \citep{Sreenivasan:POF1998}, which suggests that boundary layers do matter, despite the nearness to isotropy in the central region. 

Our DNS results are fully resolved and can thus have implications for the modeling of small-scale turbulence in coarser-grid simulation studies of mesoscale convection, particularly for the development of subgrid-scale models that go beyond the mixing length theory of \cite{Prandtl:ZAMM1925}. This class of algebraic turbulence models is still a workhorse in astrophysical simulations; see for example discussions of their limitations and extensions in \citet{Miesch:LRSP2005} and \citet{Kupka:LRCA2017}, or recent extensions by \citet{Brandenburg:APJ2016}. Finally, we stress that the convection considered here does not incorporate complexities such as rotation, magnetic field, varying molecular transport coefficients, or curvature, which are present in geophysical and astrophysical settings. They would have to be included to yield realistic models in these specific instances. The present focus has been the exploration of the effects of low Prandtl numbers, which is an important facet of these flows. A more detailed study of these points in connection with small-scale intermittency in low-Prandtl-number RBC flows is underway, and will be reported elsewhere. 

\backsection[Acknowledgements]{The authors gratefully acknowledge the Gauss Centre for Supercomputing e.V. (https://www.gauss-centre.eu) for funding this project by providing computing time on the GCS Supercomputer SuperMUC-NG at Leibniz Supercomputing Centre (https://www.lrz.de).}

\backsection[Funding]{A.P. acknowledges support from the Deutsche Forschungsgemeinschaft (DFG) within the Priority Programme ``Turbulent Superstructures'' under Grant No. DFG-SPP 1881. This work was also supported by Grant No. SCHU 1410/30-1 of DFG and NYUAD Institute Grant G1502 ``NYUAD Center for Space Science." D.K. is partly supported by Grant No. KR 4445/2-1 of DFG.} 

\backsection[Declaration of Interests]{The authors report no conflict of interest.}

\backsection[Data availability statement]{The data that support the findings of this study are available from the corresponding author upon reasonable request.}

\backsection[Author ORCIDs]{\\
A. Pandey, https://orcid.org/0000-0001-8232-6626; \\
D. Krasnov, https://orcid.org/0000-0002-8339-7749; \\
K.R. Sreenivasan, https://orcid.org/0000-0002-3943-6827; \\
J. Schumacher, https://orcid.org/0000-0002-1359-4536}

\appendix

\section{Comparison of the spectral element and finite difference solvers} \label{sec:app1}
\begin{figure}
\includegraphics[width=1\textwidth]{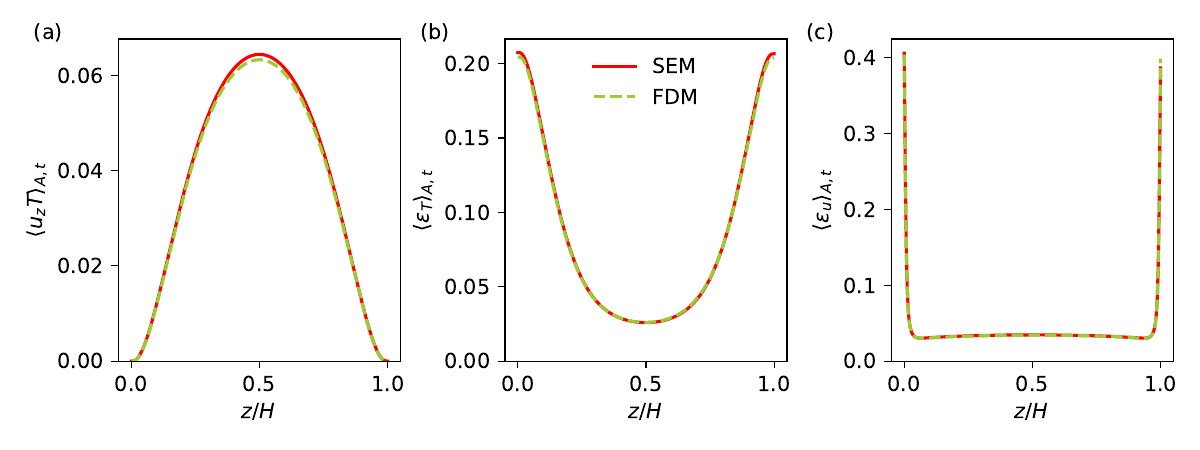}
\caption{Vertical profiles of (a) the convective heat flux, (b) the thermal dissipation rate, and (c) the kinetic energy dissipation rate computed using the spectral element (solid red curves) and the finite difference (dashed green curves) solvers for $Pr = 0.005, Ra = 10^5$ agree very well each other.}
\label{fig:comp}
\end{figure}

As mentioned in \S~\ref{sec:numerical}, we use two different solvers, one based on the spectral element method (SEM) and one based on the finite difference method (FDM), to perform our simulations due to the demanding requirements at very low Prandtl numbers. Therefore, it is important to ensure that both the solvers yield the same results. To check this, we performed comparison simulations, for $Pr = 0.005$ and $Ra = 10^5$ and for $Pr=0.7$ and $Ra=10^7$. In figure~\ref{fig:comp}, we plot the vertical profiles of the convective heat flux, the thermal, and the kinetic energy dissipation rates for the former comparison. The profiles from both the spectral element and finite difference solvers agree very well with each other. The globally-averaged quantities such as the Nusselt and the Reynolds numbers also agree excellently.

The global heat flux can also be estimated using the global dissipation rates from the exact relations~\citep{Howard:ARFM1972}, and their concurrence is also an indicator of the sufficiency of the spatial and temporal resolutions~\citep{Pandey:APJ2021}. The exact relations yield the Nusselt numbers as 
\begin{eqnarray}
Nu_{\varepsilon_u} & = & 1 + \sqrt{RaPr} \langle \varepsilon_u \rangle_{V,t} \, , \label{eq:Nu_v} \\
Nu_{\varepsilon_T} & = & \sqrt{RaPr} \langle \varepsilon_T \rangle_{V,t}. \label{eq:Nu_T}
\end{eqnarray}
Further, the heat flux $Nu(z)$ in each horizontal plane, see \eqref{eq:Nuz} in the main text, remains a constant across the convection layer in the statistically steady state and matches with $Nu$. We thus compute the averaged heat flux at the top and bottom plates as
\begin{equation}
Nu_{\partial_z T} = -\left \langle \left( \frac{\partial T}{\partial z}\right)_{z=0,H} \right\rangle_{A,t}  \label{eq:Nu_plate}
\end{equation}
and list it along with $Nu, Nu_{\varepsilon_u}, Nu_{\varepsilon_T}$ in table~\ref{table:Nu}. The results in the table show that the agreement between all differently obtained Nusselt numbers is excellent for all the simulations.

A few words on the determination of the kinetic energy dissipation rate from the results of the FDM solver. It has been shown in \citet{Vire2009} that, for the rate of strain tensor $S_{ij}=(\partial u_i/\partial x_j + \partial u_j/\partial x_i)/2$, the identity $S_{ij} S_{ij} = -u_{i} \partial S_{ij}/\partial x_j + \partial (u_{i} S_{ij})/\partial x_j$ is not fulfilled in discrete form, which is particularly relevant to finite-difference and finite-volume methods. The direct computation of $S_{ij} S_{ij}$ tends to yield underpredicted values, especially if the discretization errors are large. For example, in case of coarse grids, more pertinent to LES however, the difference between direct computation of $S_{ij} S_{ij}$ and summation-by-parts can yield a factor of 2 to 2.5 in the buffer and logarithmic layer regions~\citep{Vire2011}. Albeit this difference scales as $\sim O(h^2)$ (with $h$ being the mesh step-size) for the $2^{\rm nd}$-order approximations, it cannot be completely neglected, even in case of finer resolutions used in DNS. Here, we cannot directly apply the summation-by-parts approach, since it involves the so-called flux-variables, prescribed at the cell interface~\citep{Vire2011}. These flux-variables are used in the algorithm to secure divergence-free condition and conservative form of the non-linear terms, but not stored. We have applied a different approach for the FDM results -- the $S_{ij}$ tensor is computed by using $6^{\rm th}$-order accurate stencils for the velocity gradients to reduce the effect of discretization errors.

\begin{table}
  \begin{center}
\def~{\hphantom{0}}
  \begin{tabular}{lccccccc}
$Pr$ & $Ra$ & No. of mesh cells  & $Nu$ & $Nu_{\varepsilon_u}$ & $Nu_{\varepsilon_T}$ & $Nu_{\partial_z T}$ & $t_\mathrm{sim} (t_f)$ \\
0.001 & $10^5$ & $9600 \times 9600  \times 640$  & $1.21 \pm 0.003$ & $1.22 \pm 0.001$ & $1.22 \pm 0.003$ & $1.22 \pm 0.003$ & 5.4 \\
0.005 & $10^5$ & $2367488 \times 11^3$  & $1.86 \pm 0.02$ & $1.85 \pm 0.02$ & $1.86 \pm 0.02$ & $1.86 \pm 0.02$ & 60 \\
0.005 & $10^5$ & $8192 \times 8192 \times 512$  & $1.84 \pm 0.001$ & $1.84 \pm 0.002$ & $1.84 \pm 0.001$ & $1.84 \pm 0.001$ & 3.5 \\
0.021 & $10^5$ & $2367488 \times 7^3$  & $2.62 \pm 0.02$ & $2.62 \pm 0.02$ & $2.62 \pm 0.02$ & $2.62 \pm 0.02$ & 156 \\
0.7 & $10^5$ & $1352000 \times 5^3$  & $4.26 \pm 0.03$ & $4.26 \pm 0.03$ & $4.26 \pm 0.03$ & $4.26 \pm 0.02$ & 1108 \\
7 & $10^5$ & $1352000 \times 5^3$& $4.14 \pm 0.02$ & $4.14 \pm 0.02$ & $4.14 \pm 0.02$ & $4.14 \pm 0.01$ & 2670 \\ 
0.001 & $10^6$ & $12800  \times 12800 \times 800$ & $2.48 \pm 0.003$ & $2.47 \pm 0.02$ & $2.48 \pm 0.01$ & $2.48 \pm 0.01$ & 2.0 \\
0.005 & $10^6$ & $8192  \times 8192 \times 512$  & $3.52 \pm 0.02$ & $3.51 \pm 0.03$ & $3.52 \pm 0.02$ & $3.52 \pm 0.02$ & 6.3 \\
0.021 & $10^6$ & $8192  \times 8192 \times 512$  & $4.84 \pm 0.01$ & $4.81 \pm 0.02$ & $4.81 \pm 0.03$ & $4.81 \pm 0.02$ & 6.0 \\
0.7 & $10^6$ & $2367488 \times 7^3$ & $8.10  \pm 0.06$ & $8.10 \pm 0.04$ & $8.10 \pm 0.04$ & $8.10 \pm 0.03$ & 1292 \\
7 & $10^6$ & $2367488 \times 7^3$ & $8.30 \pm 0.05$ & $8.30 \pm 0.04$ & $8.30 \pm 0.02$ & $8.30 \pm 0.02$ & 3255 \\ 
0.001 & $10^7$ & $20480 \times 20480 \times 1280$ & $4.57 \pm 0.01$ & $4.56 \pm 0.10$ & $4.59 \pm 0.01$ & $4.59 \pm 0.01$ & 1.0 \\
0.7 & $10^7$ & $2367488 \times 11^3$ & $15.48  \pm 0.13 $ & $15.48 \pm 0.08$ & $15.48 \pm 0.06$ & $15.48 \pm 0.10$ & 248 \\
0.7 & $10^7$ & $6720 \times 6720 \times 420$ & $15.50  \pm 0.10 $ & $15.57 \pm 0.03$ & $15.52 \pm 0.03$ & $15.52 \pm 0.03$ & 10 \\
7 & $10^7$ & $1352000 \times 11^3$ & $16.25 \pm 0.08$ & $16.25 \pm 0.05$ & $16.25 \pm 0.03$ & $16.25 \pm 0.02$ & 471
  \end{tabular}
  \caption{The turbulent heat flux, which is computed in four different ways, agrees very well for all reported simulations. The error bars indicate the standard deviation. $t_\mathrm{sim}$ is the total simulation time in the statistically steady state. The time advancement by one free-fall time unit for the biggest simulation took 30 million core hours on 144000 processor cores on the cluster SuperMUC-NG at Leibniz Rechenzentrum Garching.}
  \label{table:Nu}
  \end{center}
\end{table}

\section{Grid sensitivity for our biggest simulation}
\label{sec:app2}

The Kolmogorov length scale in the flow for $Pr = 0.001$ and $Ra = 10^7$ becomes very small and, consequently, computational resources required for the numerical investigation of this flow become exorbitant. Therefore, to determine the optimum number of nodes needed to resolve the flow adequately, we performed this simulation on three different grids with $15360 \times 15360 \times 1024$, $20480 \times 20480 \times 1280$, and $22400 \times 22400 \times 1400$ mesh cells, which we denote in the following as mesh-1, mesh-2, and mesh-3, respectively. As summarized in \citet{Scheel:NJP2013}, we compare the horizontally as well as the globally averaged quantities from these simulations to test the effects of grid resolution. The kinetic energy dissipation rate field $\varepsilon_u ({\bm x},t)$, which involves the computation of all the nine terms of the velocity gradient tensor, is very sensitive to the mesh size in a low-$Pr$ convection. We computed the horizontally-averaged kinetic energy dissipation rate $\langle\varepsilon_u\rangle_{A,t}(z)$ and investigated the region near the midplane where the computational grid is coarsest. We observed underresolved data for mesh-1, whereas the variation of $\langle\varepsilon_u\rangle_{A,t}(z)$ is smooth for simulations with mesh-2 and mesh-3. Furthermore, $\langle\varepsilon_u\rangle_{A,t}(z)$ from mesh-2 and mesh-3 agree very well, in particular they both yield the same relative difference between the results of $2^{\rm nd}$- and $6^{\rm th}$-order stencils applied for the direct computation of the velocity gradients. This analysis clearly indicates that mesh-2 is able to properly capture the velocity derivatives for these extreme parameters. Thus, we carried out our simulation for $Pr = 0.001$ and $Ra = 10^7$ with mesh-2.

\section{Estimation of the characteristic lengths and times of superstructures}

\begin{figure}
\centerline{\includegraphics[width=\textwidth]{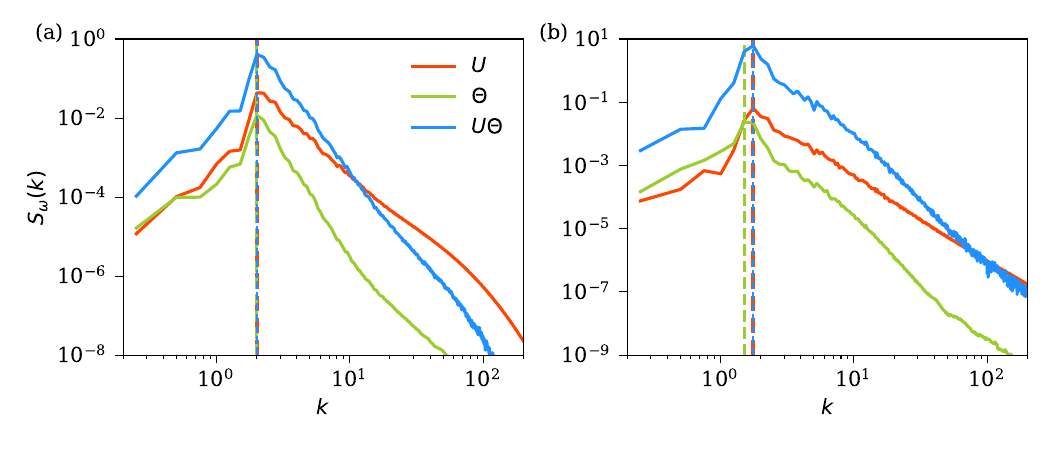}}
\caption{Power spectra $S_U$, $S_{\Theta}$, and $S_{U\Theta}$ as defined in equations~\eqref{su}--\eqref{sut} in the midplane for $Pr = 0.001$ at (a) $Ra = 10^5$ and (b) $Ra = 10^7$. They exhibit peak at nearly the same wavenumber corresponding to the characteristic spatial scale (or wavelength) of the turbulent superstructures of convection.}
\label{fig:E_k_pr0.005}
\end{figure}

We show the power spectra $S_U(k), S_\Theta(k)$, and $S_{U\Theta}(k)$ for $Pr = 0.001$ at $Ra = 10^5$ and $Ra = 10^7$ in figure~\ref{fig:E_k_pr0.005} and find that the spectral distributions for the velocity and temperature fields are different. Figure~\ref{fig:E_k_pr0.005} shows that the power first increases with decreasing length scales and attains a maximum before declining sharply with further decrease in the scale size. We find that the decay of the thermal variance spectrum beyond $k_\mathrm{max}$ is rapid compared to that of the squared vertical velocity component. This is because the velocity field in very-low-$Pr$ convection is vigorously turbulent and possesses larger fine-scale contributions compared to the predominantly large-scale nature of the temperature field~\citep{Schumacher:PNAS2015}. We find, however, that the three spectra exhibit a peak at nearly the same wavenumber $k^\omega_\mathrm{max}$ corresponding to the maximum of $S_\omega(k)$, yielding the characteristic spatial scale $\lambda_\omega = 2\pi/k^\omega_\mathrm{max}$ of the superstructures, with $\omega = \lbrace U, \Theta, U\Theta \rbrace$. 

\begin{figure}
\centerline{\includegraphics[width=0.9\textwidth]{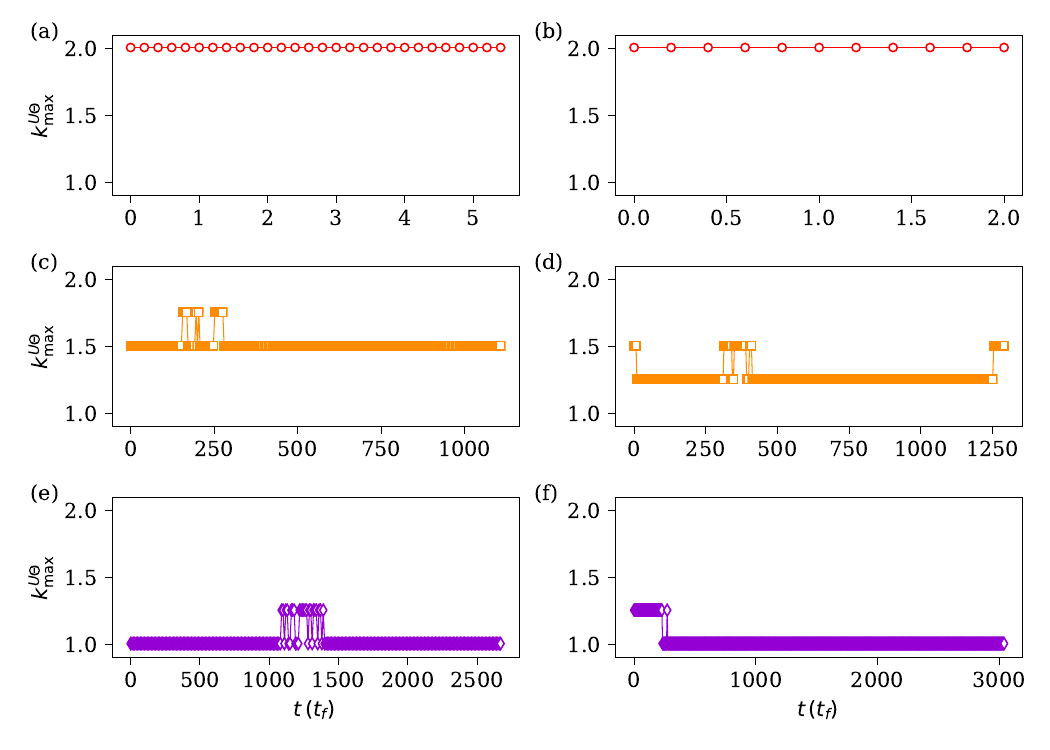}}
\caption{Temporal evolution of the peak wavenumber $k_\mathrm{max}$ for $Ra = 10^5$ (a,c,e) and $Ra = 10^6$ (b,d,f). The location of maximum does not change for $Pr = 0.001$ (a,b), whereas it occasionally shifts towards the higher wavenumber for $Pr = 0.7$ (c,d) and also for $Pr = 7$ (e,f).}
\label{fig:kmax_time}
\end{figure}

Furthermore, the spatial scale does not remain fixed but fluctuates during the evolution of the flow. This can be seen in figure~\ref{fig:kmax_time}, where we plot $k_\mathrm{max}^{U\Theta}(t)$, extracted from each instantaneous snapshot of our simulations, as a function of time.
The figure shows that $k^{U\Theta}_\mathrm{max}$ is independent of time for the entire duration of simulations for $Pr = 0.001$. The same comment also holds for $Pr = 0.005$ and $Pr = 0.021$ simulations (not shown in the figure). However, $k^{U\Theta}_\mathrm{max}$ for the simulations at $Pr = 0.7$ and $Pr = 7$, though remaining fixed for most of the time, shows occasional excursions. Thus, we determine the characteristic spatial scale of superstructures by time averaging $k^{U\Theta}_\mathrm{max}(t)$ and finding $\lambda = 2\pi/ \langle k^{U\Theta}_\mathrm{max}(t) \rangle_t$. 


\end{document}